\documentclass[fleqn,usenatbib]{mnras}

\usepackage{newtxtext,newtxmath}

\usepackage[T1]{fontenc}

\DeclareRobustCommand{\VAN}[3]{#2}
\let\VANthebibliography\thebibliography
\def\thebibliography{\DeclareRobustCommand{\VAN}[3]{##3}\VANthebibliography}

\usepackage{graphicx}	
\usepackage{amsmath}	

\usepackage{rotating}
\usepackage{multirow}
\usepackage{color}

\newcommand{\three}{\textsc{The Three Hundred} project}
\newcommand{\rs}{$r_\mathrm{s}$}
\newcommand{\ms}{$M_\mathrm{s}$}
\newcommand{\tx}{$T_\mathrm{X}$}

\newcommand{\relaxmath}{\chi_\mathrm{DS}}
\newcommand{\relaxtext}{$\relaxmath$}

\graphicspath{{./}{figures/}}

\title[The fundamental plane of galaxy clusters]{The Three Hundred Project: dissecting the fundamental plane of galaxy clusters up to $z=1$}

\author[L. A. D\'iaz-Garc\'ia et al.]{
Luis A. D\'iaz-Garc\'ia,$^{1,2}$\thanks{E-mail: luis@iaa.es (LADG)}
Keiichi Umetsu,$^{1}$
Elena Rasia,$^{3,4}$
Weiguang Cui,$^{5}$
and Massimo Meneghetti$^{6,7}$
\\
$^{1}$Academia Sinica Institute of Astronomy and Astrophysics (ASIAA), No.~1, Section~4, Roosevelt Road, Taipei 10617, Taiwan\\
$^{2}$Instituto de Astrof\'isica de Andaluc\'ia (IAA-CSIC), P.O.~Box 3004, 18080 Granada, Spain\\
$^{3}$INAF - Osservatorio Astronomico Trieste, via Tiepolo 11, 34123, Trieste 34123, Italy\\
$^{4}$Institute of Fundamental Physics of the Universe, via Beirut 2, 34151 Grignano, Trieste, Italy\\
$^{5}$Institute for Astronomy, University of Edinburgh, Edinburgh EH9 3HJ, UK\\
$^{6}$INAF - Osservatorio Astronomico di Bologna, via Ranzani 1, 40127 Bologna, Italy\\
$^{7}$INFN, Sezione di Bologna, viale Berti Pichat 6/2, I-40127 Bologna, Italy
}

\date{Accepted XXX. Received YYY; in original form ZZZ}

\pubyear{2021}

\begin{document}
\label{firstpage}
\pagerange{\pageref{firstpage}--\pageref{lastpage}}
\maketitle

\begin{abstract}

We perform a systematic study of the recently discovered fundamental plane of galaxy clusters (CFP) using $\sim250$ simulated clusters from \three, focusing on the stability of the plane against different temperature definitions and its dependence on the dynamical relaxation state of clusters. The CFP is characterised by $T\propto M_\mathrm{s}^\alpha\,r_\mathrm{s}^\beta$, defined with the gas temperature ($T$) and the characteristic halo scale radius and mass ($r_\mathrm{s}$ and $M_\mathrm{s}$) assuming a Navarro-Frenk-White halo description. We explore two definitions of weighted temperatures, namely mass-weighted and spectroscopic-like temperatures, in three radial ranges. \three\ clusters at $z=0$ lie on a thin plane whose parameters ($\alpha,\beta$) and dispersion ($0.015$--$0.030$~dex) depend on the gas temperature definition. The CFP for mass-weighted temperatures is closer to the virial equilibrium expectation ($\alpha=1,\beta=-1$) with a smaller dispersion. For gas temperatures measured within $500h^{-1}$~kpc, the resulting CFP deviates the most from the virial expectation and shifts towards the similarity solution for a secondary infall model ($\alpha=1.5,\beta=-2$). Independently of the temperature definition, we find that clusters at $z=1$ and relaxed clusters form a CFP similar to the virial expectation, unlike disturbed clusters exhibiting stronger evolution. Only systems formed over the last 4~Gyr present a CFP that is closer to the similarity solution. All these findings are compatible with the CFP obtained for a CLASH subsample excluding the hottest clusters with $T_\mathrm{X}>12$~keV.

\end{abstract}



\section{Introduction} \label{sec:introduction}


Clusters of galaxies are the largest class of gravitationally bound objects to have formed in the universe. The mass content of galaxy clusters is proved to be dominated by elusive dark matter ($\sim 85$~per cent), while the vast majority of the baryons residing in the cluster potential well are in the hot X-ray emitting phase \citep[$\sim 10$--$13$~per cent in total mass; e.g.,][]{Vikhlinin2006,Umetsu2009,Planck2013}. Massive galaxy clusters thus contain a wealth of information about the initial conditions for cosmic structure formation and the growth of structure across cosmic time \citep[][]{Frenk1990,Blandford1992,White1993,Jenkins2001,Reiprich2002}. 
    
The distribution of dark matter in quasi equilibrium halos, such as galaxy clusters, depends fundamentally on the properties of dark matter. Even though the formation of halos is a complex and nonlinear dynamical process and they are continuously evolving through accretion and mergers, Lambda cold dark matter ($\Lambda$CDM) models predict that the average density profile $\rho(r)$ of collisionless halos in quasi gravitational equilibrium is well described by the Navarro--Frenk--White profile \citep[][hereafter NFW]{NFW1996,NFW1997} out to their virial radius. The NFW radial density profile is fully specified by two parameters and is defined as
\begin{equation}\label{eq:nfw}
    \rho(r)=\frac{\delta_\mathrm{c}\rho_\mathrm{c}}{(r/r_\mathrm{s})(1+r/r_\mathrm{s})^2},
\end{equation}
where $\rho_\mathrm{c}$ is the critical density of the universe at the halo redshift $z$, $r_\mathrm{s}$ is the characteristic scale radius at which the logarithmic density slope equals $-2$, and $\delta_\mathrm{c}$ sets the normalisation of the profile. In the present paper, we often describe the NFW model using $r_\mathrm{s}$ and the halo mass enclosed within it, $M_\mathrm{s}\equiv M(<r_\mathrm{s})$, also referred to as the characteristic scale mass.

According to $N$-body simulations of $\Lambda$CDM models, the structural parameters of the NFW profile, such as halo concentration $c_\Delta \equiv r_\Delta/r_\mathrm{s}$, are closely linked to the growth history of individual halos \citep{NFW1997,Ludlow2013}. Here $r_\Delta$ is the overdensity radius at which the mean interior density is $\Delta$ times the critical density of the universe at the specific redshift (i.e.~$\Delta\times\rho_\mathrm{c}(z)$). The inner region of halos ($r\lesssim r_\mathrm{s}$) develops in an initial phase of rapid growth, which is often associated to major mergers with other halos. During the subsequent slow accretion phase, the scale radius $r_\mathrm{s}$ stays roughly constant, whereas $r_\Delta$ continues to grow through a mixture of physical mass accretion and pseudo-evolution caused by the decrease of $\rho_\mathrm{c}$ over time \citep[e.g.,][]{Diemer2013}. Halos thus form `inside-out'. In this context, halos that formed earlier tend to present a higher characteristic density $\rho_\mathrm{s}\equiv 3M_\mathrm{s}/(4\pi r_\mathrm{s}^3)$ \citep{NFW1997,Zhao2003,Ludlow2013}. In contrast to $\rho_\mathrm{c}(z)$, this characteristic density $\rho_\mathrm{s}$ is not constant from cluster to cluster but depends on the formation time of each cluster.
    
Recently, \citet{Fujita2018a,Fujita2018b} discovered a new fundamental plane of galaxy clusters (CFP) using gravitational lensing and X-ray observations of $20$ high-mass clusters available from the Cluster Lensing And Supernova survey with Hubble \citep[][hereafter CLASH]{Postman2012,Donahue2014,Umetsu2014,Umetsu2016,Merten2015}. \citet{Fujita2018a} showed that the $20$ CLASH clusters lie on a thin plane defined in the three-dimensional logarithmic space of their characteristic scale radius $r_\mathrm{s}$, mass $M_\mathrm{s}$, and X-ray temperature $T_\mathrm{X}$ of the intracluster medium (ICM) with an orthogonal scatter of $\sim 0.045$~dex. Their findings suggest that the parameters ($M_\mathrm{s}, r_\mathrm{s}$) characterising the internal structure of dark-matter halos are tightly coupled with the gas temperature $T_\mathrm{X}$. Based on the tight correlation they found, \citet{Fujita2018a} argue that the gas temperature should reflect the potential depth of dark-matter halos at a specific cluster formation time, which is encoded in $M_\mathrm{s}$ and $r_\mathrm{s}$. Intriguingly, the plane is tilted with respect to $T_\mathrm{X}\propto M_\mathrm{s}\, r_\mathrm{s}^{-1}$, the plane expected in the case of simplified virial equilibrium. \citet{Fujita2018a} found that this tilt can be explained by a spherical similarity solution for secondary infall and accretion of gas in a matter-dominant universe \citep{Bertschinger1985}, which predicts $T_\mathrm{X}\propto M_\mathrm{s}^{1.5}\, r_\mathrm{s}^{-2}$ at the cluster scales \citep{Fujita2018a}. They also found that cosmological $N$-body/hydrodynamical simulations reproduce the observed plane and its tilt angle for cluster-scale halos.

The CFP has been proposed to describe the connection between the thermodynamic history of the intracluster gas and the evolution of the internal structure of dark-matter halos \citep{Fujita2018a}. With this framework in mind it is clear that studying the CFP as a function of redshift provides an important clue for improving our understanding of how galaxy clusters form and evolve through mergers and accretion. The pioneering works of \citet{Fujita2018a,Fujita2018b}, however, did not study in detail to what extent the CFP parameters and scatter depend on the cluster redshift, the dynamical state of clusters, the definition of the gas temperature, and the gas physics and feedback processes implemented in the code. In this paper, we supply this deficiency and present a systematic study of this fundamental plane using a sample of $\sim 250$ simulated cluster-scale halos \citep[ten times larger than that of\ ][]{Fujita2018a,Fujita2018b} modelled with full-physics hydrodynamical re-simulations by \three\ \citep{Cui2018}, focusing specifically on the redshift evolution of the plane, the dependence on the dynamical state of halos, and the stability of the plane against different temperature definitions.
    
This paper is organised as follows. In Section~\ref{sec:data}, we present the sample of simulated galaxy clusters of \three\ as well as the sample of CLASH clusters used in this research. The methodology to determine the parameters defining the CFP is described in Section~\ref{sec:method}, whereas the CFPs obtained from the analysis of the \three\ sample at $z=0$ is described in Section~\ref{sec:fplane300}. In Section~\ref{sec:cfp_z}, we study via simulations whether there is an evolution of the CFP with redshift. The dynamical state of \three\ halos and its impact on the CFP is explored in Section~\ref{sec:relaxation}. The CFP obtained for CLASH clusters by real observations is presented in Section~\ref{sec:fplaneclash}. Our results and conclusions are discussed and compared with previous results of the literature in Section~\ref{sec:discussion}. Finally, a summary of the present research is included in Section~\ref{sec:summary}. 
    
Throughout this paper, we adopt a spatially flat $\Lambda$CDM cosmology with the same parameters used in \three\ based on the Planck 2015 cosmology \citep[][]{Planck2016}, namely: Hubble constant of $H_0=67.8$~km~s$^{-1}$~Mpc$^{-1}$, $\Omega_\mathrm{M}=0.307$ (total matter density), $\Omega_\mathrm{b}=0.048$ (baryon density), $\Omega_\Lambda=0.693$ (cosmological constant density), $\sigma_8=0.823$ (matter power spectrum normalisation), and $n_\mathrm{s}=0.96$ (scalar spectral index). We use the standard notation $M_{\Delta}$ for the mass enclosed within a sphere of radius $r_{\Delta}$, within which the mean overdensity equals $\Delta \times \rho_\mathrm{c}(z)$ at a particular redshift $z$. That is, $M_{\Delta}=(4\pi\Delta/3)\rho_\mathrm{c}(z)r_{\Delta}^3$.


\section{Galaxy cluster samples}\label{sec:data}


\subsection{\three}\label{sec:300th}

\three\ \citep[][]{Cui2018} is composed of $324$ galaxy clusters with masses above $M_{200}>6.24\times 10^{14}h^{-1}~M_\odot$ and located at the centre of zoom-in resimulated regions of radius $15h^{-1}$~Mpc. These regions were selected from the fiducial $N$-body and dark-matter-only MultiDark Planck 2 simulation at redshift $z=0$ \citep[MDPL2,][]{Klypin2016}, to subsequently carry out a re-simulation with hydrodynamics physics. The MDPL2 has a size of comoving length $1$~$h^{-1}$~Gpc and is simulated with $3840^3$ dark matter particles with a mass resolution of $1.5\times 10^{9}h^{-1}$~Mpc. For the resimulation of \three, the effective mass resolutions for dark matter and gas are $m_\mathrm{DM}=1.27 \times 10^9h^{-1}~M_\odot$ and $m_\mathrm{gas}=2.36 \times 10^8h^{-1}~M_\odot$, respectively. That is, the same combined mass resolution as for MDPL2 according to a baryon fraction of $\Omega_\mathrm{b}/\Omega_\mathrm{M}\sim 0.16$ (Planck 2015 cosmology).

\three\ accounts for a broad range of baryonic physics in the \textsc{GADGET-X} code \citep[][]{Rasia2015}, which utilises a modern Smooth-Particle-Hydrodynamic (SPH) scheme based on a modified version of the \textsc{GADGET3} code \citep[for details, see][]{Beck2016}. The code implements a supermassive black hole accretion and active galactic nuclei (AGN) feedback by \citet[][]{Steinborn2015}. In addition, the \textsc{GADGET-X} code includes a metal dependent model for gas cooling \citep[][]{Wiersma2009}, an homogeneous ultra-violet background \citep[following][]{Haardt1996}, a star formation model accounting for the metal enrichment of the ICM \citep[][]{Tornatore2007} assuming a \citet[][]{Chabrier2003} stellar initial mass function, and supernova feedback \citep[][]{Springel2003}. For further details about the \textsc{GADGET-X} code and comparison with respect to other codes and schemes, we refer readers to \citet[][]{Cui2018}. It should be noted that although the resimulated Lagrangian regions contains multiple groups and clusters, we focus in this work on the central objects. With the aim of determining the redshift evolution of the CFP, we also study in detail \three\ clusters at higher redshifts. Specifically, we explore the CFP at $z=0.00, 0.07, 0.22, 0.33, 0.59$, and $0.99$.


\subsection{Galaxy clusters from the CLASH program}\label{sec:clash}

One of the main goals of the CLASH program was to precisely constrain the mass density profiles of $25$ galaxy clusters using deep lensing observations. The sample of galaxy clusters targeted by the CLASH program is subdivided into two subsamples: (i) $20$ hot X-ray clusters with $T_\mathrm{X}>5$~keV and nearly concentric X-ray isophotes, as well as a well-defined X-ray peak closely located to the BCG position; and (ii) five clusters selected by their exceptional lensing strength (characterised by large Einstein radii, $\theta_\mathrm{Ein}>35\arcsec$, for a fiducial source at redshift $z=2$) so as to magnify galaxies at high redshift.

It is worth emphasising that the X-ray subsample was not based on a lensing preselection to avoid a biased sample towards intrinsically concentrated clusters and/or those systems where the major axis is preferentially aligned with the line of sight \citep[][]{Hennawi2007,Oguri2009,Meneghetti2010}. Numerical simulations suggest that the CLASH X-ray-selected subsample is mostly (but not exclusively) composed of relaxed systems ($\sim 70$~per cent) and largely free of such orientation bias \citep[][see also Table~\ref{tab:clash}]{Meneghetti2014}. On the other hand, high-magnification-selected clusters often turn out to be dynamically disturbed as a consequence of highly massive ongoing mergers \citep[][see also references therein]{Umetsu2020rev}.

For an observational determination of the fundamental plane, we combined redshift and X-ray temperature information detailed in \citet[][]{Postman2012} along with the characteristic scale radius and mass measurements from \citet[][]{Umetsu2016}, yielding a subsample of $20$ CLASH clusters composed of $16$ X-ray-selected and $4$ high-magnification clusters (see Section~\ref{sec:method_clash} for further details). It should be noted that $5$ of the $25$ clusters of the CLASH sample were not included in the joint lensing analysis performed by \citet{Umetsu2016} because they lacked of usable wide-field ground-based weak-lensing data \citep[see][]{Umetsu2014}. Consequently, they were also excluded in this work. The CLASH subsample spans a redshift range of $0.19 \le z \le 0.69$ with a median redshift of $0.35$. \citet{Umetsu2016} found that the stacked strong- and weak-lensing signal of the CLASH X-ray-selected subsample is best described by the NFW model. In Table~\ref{tab:clash}, we summarise the main properties of the CLASH clusters used in this work. As a consequence of the CLASH selection, all the clusters show X-ray temperatures higher than $5$~keV. 

\begin{figure*}
\centering
\includegraphics[width=0.8\textwidth,clip=True]{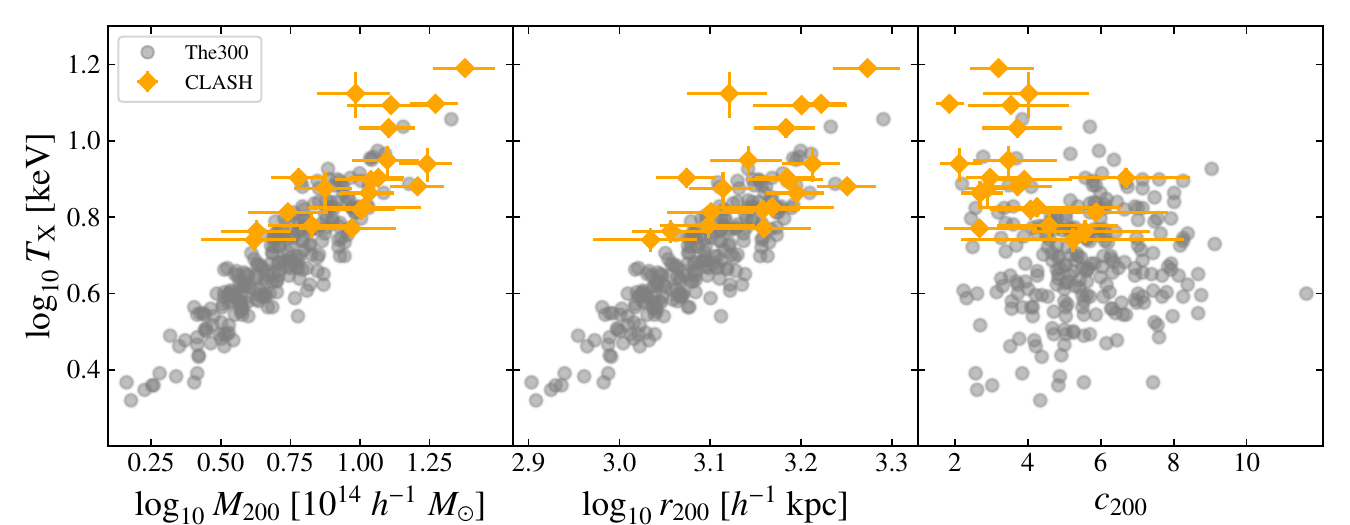}
\caption{Mass, radius, and concentration (horizontal axis, from left to right) versus X-ray temperature (vertical axis, $T_\mathrm{X}$ extracted in the radial range $50$--$500 h^{-1}$~kpc) of the CLASH cluster sample (at a median redshift of $z=0.36$; orange diamonds) and simulated galaxy clusters of \three\ (at $z=0.33$; grey circles) for spectroscopic-like temperatures (vertical axis, $T_\mathrm{sl}$ measured within the range $50$--$500h^{-1}$~kpc). Vertical and horizontal bars show the $1\sigma$ errors of parameters for CLASH clusters.
\label{fig:comparison_sample}}
\end{figure*}

\begin{table*}

\centering
\caption{Properties of galaxy clusters selected from CLASH. All halo mass and radius measurements were obtained from a joint strong lensing, weak lensing shear-and-magnification analysis of \citet[][]{Umetsu2016}, whereas the rest of properties were taken from \citet[][]{Postman2012}.}
\label{tab:clash}

\begin{tabular}{ccccccccc}
\hline
Cluster & RA & DEC & \textit{z} & $M_{200}$ & $r_{200}$ & $M_\mathrm{s}$ & $r_\mathrm{s}$ & $T_\mathrm{X}$ \\
 & [deg] & [deg] & & [$10^{14}h^{-1}M_\odot$] &  [$h^{-1}$~kpc]  & [$10^{14}h^{-1}M_\odot$] &  [$h^{-1}$~kpc] & [keV] \\
\hline
X-Ray-selected &&&&&&&& \\
Abell 383 & $\ 42.014$ & $-\ 3.529$ & $0.187$ & $\ 5.51^{+2.16}_{-1.56}$ & $1261^{+147}_{-132}$ & $0.99^{+0.70}_{-0.39}$ & $215^{+111}_{-69}$ & $6.5\pm0.24$ \\
Abell 209$^*$ & $\ 22.969$ & $-13.611$ & $0.206$ & $10.76^{+2.49}_{-2.26}$ & $1567^{+112}_{-118}$ & $3.64^{+1.50}_{-1.12}$ & $584^{+167}_{-133}$ & $7.3\pm0.54$ \\
Abell 2261$^*$ & $260.614$ & $+32.133$ & $0.224$ & $16.11^{+3.88}_{-3.31}$ & $1782^{+133}_{-132}$ & $4.10^{+1.91}_{-1.30}$ & $480^{+163}_{-120}$ & $7.6\pm0.30$ \\
RX~J2129.7$+$0005 & $322.416$ & $+\ 0.089$ & $0.234$ & $\ 4.25^{+1.40}_{-1.10}$ & $1139^{+113}_{-108}$ & $0.80^{+0.48}_{-0.29}$ & $205^{+93\ }_{-61\ }$ & $5.8\pm0.40$ \\
Abell 611 & $120.237$ & $+36.057$ & $0.288$ & $10.96^{+3.36}_{-2.85}$ & $1534^{+143}_{-146}$ & $2.69^{+1.63}_{-1.04}$ & $395^{+174}_{-123}$ & $7.9\pm0.35$ \\
MS 2137$-$2353 & $325.063$ & $-23.661$ & $0.313$ & $\ 9.36^{+4.08}_{-3.32}$ & $1443^{+185}_{-196}$ & $3.22^{+3.62}_{-1.79}$ & $543^{+392}_{-244}$ & $5.9\pm0.30$ \\
RX~J2248.7$-$4431$^*$ & $342.185$ & $-44.530$ & $0.348$ & $12.93^{+5.40}_{-3.97}$ & $1587^{+196}_{-183}$ & $3.43^{+3.28}_{-1.60}$ & $449^{+292}_{-170}$ & $12.4\pm0.60$ \\
MACS~J1115.9$+$0129 & $168.967$ & $+\ 1.499$ & $0.352$ & $11.64^{+2.79}_{-2.53}$ & $1530^{+114}_{-120}$ & $3.61^{+1.71}_{-1.20}$ & $518^{+179}_{-135}$ & $8.0\pm0.40$ \\
MACS~J1931.8$-$2635 & $292.957$ & $-26.576$ & $0.352$ & $10.39^{+6.21}_{-3.89}$ & $1473^{+249}_{-213}$ & $2.38^{+3.16}_{-1.26}$ & $347^{+300}_{-152}$ & $6.7\pm0.40$ \\
RX~J1532.9$+$3021 & $233.224$ & $+30.350$ & $0.363$ & $\ 4.17^{+1.74}_{-1.48}$ & $1083^{+134}_{-147}$ & $0.85^{+1.13}_{-0.35}$ & $207^{+304}_{-80}$ & $5.5\pm0.40$ \\
MACS~J1720.3$+$3536 & $260.071$ & $+35.607$ & $0.391$ & $10.09^{+3.28}_{-2.69}$ & $1439^{+142}_{-141}$ & $2.39^{+1.58}_{-0.94}$ & $354^{+171}_{-112}$ & $6.6\pm0.40$ \\
MACS~J0429.6$-$0253 & $\ 67.400$ & $-\ 2.886$ & $0.399$ & $\ 6.72^{+2.84}_{-2.06}$ & $1253^{+156}_{-144}$ & $1.45^{+1.24}_{-0.64}$ & $274^{+164}_{-99}$ & $6.0\pm0.44$ \\
MACS~J1206.2$-$0847$^*$ & $181.551$ & $-\ 8.801$ & $0.440$ & $12.68^{+3.07}_{-2.77}$ & $1526^{+114}_{-120}$ & $3.24^{+1.69}_{-1.12}$ & $412^{+169}_{-121}$ & $10.8\pm0.60$ \\
MACS~J0329.7$-$0211$^*$ & $\ 52.424$ & $-\ 2.197$ & $0.450$ & $\ 6.02^{+1.49}_{-1.24}$ & $1186^{+91\ }_{-88\ }$ & $1.00^{+0.43}_{-0.28}$ & $177^{+65\ }_{-44\ }$ & $8.0\pm0.50$ \\
RX~J1347.5$-$1145$^*$ & $206.877$ & $-11.753$ & $0.451$ & $23.85^{+6.60}_{-5.57}$ & $1876^{+159}_{-159}$ & $6.90^{+3.91}_{-2.50}$ & $588^{+236}_{-167}$ & $15.5\pm0.60$ \\
MACS~J0744.9$+$3927$^*$ & $116.220$ & $+39.457$ & $0.686$ & $12.53^{+3.72}_{-3.16}$ & $1386^{+126}_{-128}$ & $3.40^{+2.17}_{-1.36}$ & $401^{+190}_{-133}$ & $8.9\pm0.80$ \\
\hline
High-magnification-selected &&&&&&&& \\
MACS~J0416.1$-$2403 & $\ 64.039$ & $-24.068$ & $0.396$ & $\ 7.49^{+1.90}_{-1.69}$ & $1301^{+102}_{-106}$ & $2.37^{+1.07}_{-0.77}$ & $450^{+143}_{-109}$ & $7.5\pm0.80$ \\
MACS~J1149.5$+$2223 & $177.399$ & $+22.399$ & $0.544$ & $17.43^{+4.06}_{-3.63}$ & $1632^{+118}_{-122}$ & $7.50^{+3.74}_{-2.55}$ & $776^{+281}_{-202}$ & $8.7\pm0.90$ \\
MACS~J0717.5$+$3745 & $109.382$ & $+37.755$ & $0.548$ & $18.68^{+3.89}_{-3.54}$ & $1668^{+108}_{-113}$ & $9.19^{+3.65}_{-2.67}$ & $907^{+240}_{-188}$ & $12.5\pm0.70$ \\
MACS~J0647.7$+$7015 & $101.958$ & $+70.247$ & $0.584$ & $\ 9.66^{+3.20}_{-2.67}$ & $1321^{+132}_{-135}$ & $2.32^{+1.64}_{-0.96}$ & $329^{+178}_{-114}$ & $13.3\pm1.80$ \\
\hline

\multicolumn{9}{l}{\footnotesize
$^{*}$ Clusters with any substructure according to reports based on X-ray morphology \citep[][]{Schmidt2007,Maughan2008,Postman2012}.}\\

\end{tabular}
\end{table*}

In Fig.~\ref{fig:comparison_sample}, we show the distributions of $M_{200}$, $r_{200}$, $c_{200}$, and $T_\mathrm{X}$ (obtained in the radial range $50$--$500h^{-1}$~kpc, further details in Section~\ref{sec:method_clash}) for both observed and simulated clusters studied in this work. As a consequence of the CLASH preselection of $T_\mathrm{X} > 5$~keV, we see that \three\ includes clusters with spectroscopic-like temperatures much lower than the CLASH sample (see the left and middle panels in Fig.~\ref{fig:comparison_sample}). On the other hand, the radius and mass distributions of the \three\ sample at $T_\mathrm{sl}>5$~keV properly match the ranging values of the CLASH sample. In terms of concentration, the values for the simulated $T_\mathrm{sl}>5$~keV clusters span the range from $\sim 2$ to $\sim 9$, while all the CLASH clusters present concentrations of $c_{200}\lesssim 8$ within the $1\sigma$ uncertainty level. The \three\ sample thus includes clusters with slightly higher concentrations than estimated for the $20$ CLASH clusters. We refer to \citet{Merten2015} and \citet{Umetsu2016} for detailed studies of the CLASH concentration--mass relation and \citet{Meneghetti2014} for its theoretical prediction. Although the CLASH sample is composed of X-ray and lensing-selected clusters, in contrast to the mas-selected \three\ sample, we find a proper overlap of both samples in the parameter space (see Fig.~\ref{fig:comparison_sample}).

Finally, it is not possible to verify whether the two samples have similar relaxation states and formation times, because the dynamical states of individual CLASH clusters are uncertain. Nevertheless, the fraction of relaxed clusters in the \three\ sample \citep[typically below $40$~per cent; see][]{Cui2018} is lower than predicted for the CLASH sample ($\sim 70$~per cent). In this context, the CLASH sample may not be faithfully representative of the full cluster population with $T_\mathrm{X}>5$~keV.


\section{Constraints on the fundamental plane of galaxy clusters: methodology} \label{sec:method}


A detailed dissection of the CFP involves a set of cluster parameters that need to be constrained in a reliable manner. For simulated clusters, the characteristic scale parameters can be directly obtained from an NFW fit to the three-dimensional mass distribution (see Section~\ref{sec:method_NFW}), and we can compute the global gas temperature of the ICM by choosing an appropriate weighting scheme (see Section~\ref{sec:method_temperature}). Moreover, we can explore the influence of the dynamical relaxation state of each individual cluster on the definition of the CFP (see Section~\ref{sec:method_dynamical}). For CLASH clusters, we employ measurements of the characteristic scale parameters and the X-ray gas temperature from the published literature (see Section~\ref{sec:method_clash}), which are subject to observational uncertainties and systematics. Finally, various methods to determine the CFP are considered and the minimum distance method has been selected as the most reliable one (Section~\ref{sec:method_eq}). 


\subsection{NFW fitting of simulated clusters}\label{sec:method_NFW}

From \three\ hydrodynamical simulations, we extract the total mass profiles $M(<r)$ of individual clusters considering all particle species (dark matter, gas, stars) in three-dimensional spherical shells. Subsequently, we fit the three-dimensional mass profiles with a corresponding parametric NFW mass description. For the minimisation procedure, we use the Levenberg--Marquardt algorithm implemented in the \texttt{MPFITFUN}\footnote{\url{https://www.harrisgeospatial.com/docs/mpfitfun.html}} routine of \texttt{IDL}\footnote{\url{https://www.harrisgeospatial.com/Software-Technology/IDL}}. The fitting procedure, in addition to the normalisation, returns the scale radius, which we use to compute the characteristic mass, \ms.

For the fitting procedure of \three\ clusters at $z=0$, we fit the NFW formula to the mass profiles with the radial range $[0.08, 1.0]\times r_{100}$. For the cluster simulations at higher redshifts ($z=0.07$, $0.22$, $0.33$, $0.59$, and $0.99$), the NFW fit was performed in the radial range $[0.1, 1.0]\times r_{200}$. The innermost parts of clusters, or cores, were excised during the determination of the characteristic parameters to not be affected by the numerical resolution of our sample. This excision is common in literature since the interactions in the inner bins are dominated by two-body collisions \citep[][]{Mostoghiu2019} and the density profile at $r < 7h^{-1}$~kpc can deviate from an NFW profile becoming steeper for the presence of the brightest cluster galaxy (BCG), which may dominate the mass profile with respect to a dark-matter-only NFW profile \citep[][]{Schaller2015}. 

We exclude from the following analysis all clusters non properly described by an NFW profile, meaning that we use only objects with $\sigma_\mathrm{fit} \equiv \chi^2/\mathrm{dof} \le 2$ (average sum of the squared residuals per degree of freedom) and we exclude the systems whose scale radius is equal to the lower and upper bounds imposed during the minimisation process ($0.01$ and $1.0h^{-1}$~Mpc). These two criteria were applied for all NFW profiles at any redshift to facilitate the interpretation and understanding of this fundamental relation (details in Section~\ref{sec:cfp_z}). The number of clusters in our sample at $z=0$ amounts to $\sim245$, whereas at higher redshift this value ranges from $231$ to $280$ clusters (more details in Section~\ref{sec:cfp_z}).


\subsection{Temperature measurements of simulated clusters}\label{sec:method_temperature}

As a result of the two-phase accretion, typical of galaxy clusters, the formation time and evolution of these massive halos is encoded in the thermodynamic history of the diffuse hot gas residing in the cluster potential well. Indeed, the largest variation in gas temperature happens as a consequence of major mergers and somehow reflects the assembly of the internal structure of cluster-scale halos.

We compute gas temperatures for a set of radial ranges to explore the impact of this choice on the resulting parameters defining the fundamental plane, as well as their implications. In the present work, we compute average temperatures in radial ranges of $[0.1, 1.0]\times r_{200}$, $[0.15, 1.0]\times r_{500}$, and $[50, 500]\times h^{-1}$~kpc. These gas temperatures are denoted as $T(r_{200})$, $T(r_{500})$, and $T(500h^{-1}~\mathrm{kpc})$, respectively.

In addition, two definitions of weighted temperatures were used: mass-weighted and spectroscopic-like temperatures. For the former, the weighting function is the associated mass for each gas element in the ICM \citep[see e.g.][]{Kang1994,Bartelmann1996,Mathiesen2001} and it is more physically motivated. On the other hand, the spectroscopic-like temperature better approximates spectroscopic temperatures from X-ray observations \citep[][]{Mazzotta2004}. Formally, these temperatures are respectively expressed as follows:
\begin{equation}\label{eq:temperature_mw}
T_\mathrm{mw}= \frac{\int{m\ T~\mathrm{d}V}}{\int{m~\mathrm{d}V}} = \sum_i{m_i\ T_i}/\sum_i{m_i},
\end{equation}
\begin{equation}\label{eq:temperature_sl}
T_\mathrm{sl}= \frac{\int{n^2 T^{1/4}~\mathrm{d}V}}{\int{n^2 T^{-3/4}~\mathrm{d}V}} = \sum_i{n_i^2 T_i^{1/4}}/\sum_i{n_i^2 T_i^{-3/4}},
\end{equation}
where $m_i$ is the mass of the \textit{i}$^\mathrm{th}$ gas element, $n_i$ is the gas density, and $T_i$ is its temperature. We note that spectroscopic-like temperatures serve as a useful proxy for X-ray temperatures obtained from real observations \citep[][]{Mazzotta2004}, which also facilitates any interpretation of observational results against simulations.


\subsection{Dynamical state parameters of simulated clusters}\label{sec:method_dynamical}

Within the simulated Lagrangian regions, halos and sub-halos were identified with the AHF code \citep[][]{Knollmann2009}. These catalogues were used to compute different indicators for the cluster dynamical relaxation. The first indicator, derived from the merger tree of each halo, is the formation redshift, $z_\mathrm{form}$, defined in \citet{Mostoghiu2019} as the redshift at which half of the mass $M_{200}$ at present was already accreted, meaning that
\begin{equation}\label{eq:zform}
z_\mathrm{form} = -\ln(0.5)/\lambda\ ,
\end{equation}
where $\lambda$ is the mass accretion rate of the halo and the accretion history of each cluster is assumed as the functional form proposed by \citet{Wechsler2002} or equivalently 
\begin{equation}\label{eq:Mzform}
M_{200}(z_\mathrm{form}) = M_{200}^{z=0}\ \exp\left(-\lambda\,z\right)\ .
\end{equation}

The other dynamical relaxation indicators are the virial ratio ($\eta$), the centre-of-mass offset ($\Delta_\mathrm{r}$), and the fraction of mass in sub-halos \citep[$f_\mathrm{s}$; see][]{Cui2018}. In brief, $\eta$ relies on the kinetic, potential energy and surface pressure to determine the dynamical state of a cluster \citep[][]{Cui2017}, $\Delta_\mathrm{r}$ on the deviation of the centre-of-mass with respect to the maximum density peak of the halo, and $f_\mathrm{s}$ on the fraction of mass in subhaloes. These three indicators were measured within the radius $r_{200}$. In addition, we considered an extra indicator that combines the three indicators just mentioned \citep[see also][and further details in Section~\ref{sec:relaxation}]{Haggar2020,DeLuca2021} formally expressed as

\begin{equation}\label{eq:relax}
\relaxmath = \left[ \frac{\left(\frac{|1-\eta|}{0.15}\right)^2 + \left(\frac{\Delta_\mathrm{r}}{0.04}\right)^2 +\left(\frac{f_\mathrm{s}}{0.1}\right)^2}{3} \right]^{-\frac{1}{2}}\ .
\end{equation}


\subsection{Fundamental plane parameters of the CLASH sample}\label{sec:method_clash}

Using high-quality gravitational lensing data available for the CLASH sample, \citet[][]{Umetsu2016} reconstructed binned surface mass density profiles for $16$ X-ray-selected and $4$ high-magnification CLASH clusters. Their lensing analysis combines wide-field shear and magnification weak-lensing constraints primarily from the \textit{Subaru} telescope \citep{Umetsu2014} and small-scale weak and strong lensing constraints from the \textit{Hubble Space Telescope} \citep{Zitrin2015}. All individual mass profiles were subsequently fitted to a projected NFW profile within $2h^{-1}$~Mpc using the full covariance matrix \citep[see][]{Umetsu2016,Umetsu2020rev}. The NFW characteristic parameters \rs\ and \ms\ were determined from the posterior distributions of the NFW parameters $M_{200}$ and $c_{200}$ \citep[for further details, see][]{Umetsu2016}. It should be noted that the \rs\ and \ms\ parameters for \three\ are obtained from a three-dimensional analysis, whereas the NFW parameters for the CLASH clusters were derived from the projected lensing profiles assuming a spherical NFW halo. The effect of scatter due to triaxial halo shapes was properly accounted for in the covariance matrix for each individual cluster \citep{Umetsu2016}. On the other hand, as discussed earlier (see Section~\ref{sec:clash}), the CLASH sample should not be affected by orientation bias, so that the NFW parameters estimated for the CLASH sample are not expected to be significantly affected by the projection effects.

All CLASH clusters have observations from the \textit{Chandra X-ray Observatory} and derived X-ray properties \citep[][]{Postman2012,Donahue2014}. In particular, the X-ray temperatures presented in \citet[][]{Postman2012} were obtained using core-excised \textit{Chandra} spectra within a uniform radial range extending from $50 h^{-1}$ to $500 h^{-1}$~kpc. The X-ray temperatures obtained in this way are unaffected by the presence of a cool core \citep[see e.g.][]{Markevitch1998}. For the cosmological parameters used in \three, this radial range corresponds to $73.75$ and $737.5$~kpc, respectively.  


\subsection{Determination of the fundamental plane of clusters}\label{sec:method_eq}

The CFP is expressed by an equation of the form $T \propto M_\mathrm{s}^\alpha r_\mathrm{s}^\beta$, where the coefficients $\alpha$ and $\beta$ are tightly connected to the properties of the dark-halo structure. Therefore, $\alpha$ and $\beta$ are the most relevant parameters in this work. In fact, their values can be predicted for different theoretical scenarios and/or under certain assumptions about the relaxation/thermalisation processes. For instance, in case of simplified virial equilibrium, we expect values of $\alpha=1$ and $\beta=-1$, while the similarity solution for a secondary infall model \citep[][]{Bertschinger1985,Fujita2018a,Fujita2018b} implies $\alpha=1.5$ and $\beta=-2$. To facilitate our analysis and interpretation, we characterise the CFP in the following form:
\begin{equation}\label{eq:plane}
    \log_{10}\left(\frac{T}{T_0}\right) = \alpha\ \log_{10}\left(\frac{M_\mathrm{s}}{M_\mathrm{s,0}}\right) + \beta\ \log_{10}\left(\frac{r_\mathrm{s}}{r_\mathrm{s,0}}\right) + \delta\ , 
\end{equation}
where $M_\mathrm{s,0}$, $r_\mathrm{s,0}$, and $T_\mathrm{0}$ are respectively the median values of the characteristic scale mass, radius, and temperature (mass-weighted, spectroscopic-like, or X-ray temperature) of the clusters employed to fit the CFP. As a reference, these median values correspond to $M_\mathrm{s,0}=2.03\times 10^{14}h^{-1}~M_\odot$, $r_\mathrm{s,0}=424 h^{-1}$~kpc, and $T_\mathrm{0}=5.0$--$6.7$~keV (depending on the temperature definition and radial range) for \three\ clusters (see also Section~\ref{sec:fplane300}); whereas for the CLASH clusters these are $M_\mathrm{s,0}=2.98\times 10^{14}h^{-1}~M_\odot$, $r_\mathrm{s,0}=407 h^{-1}$~kpc, and $T_\mathrm{X,0}=7.8$~keV (see also Section~\ref{sec:fplaneclash}). In equation~(\ref{eq:plane}), there is another parameter denoted as $\delta$, which is the normalisation of the CFP. When $\delta = 0$, the plane is centred at the sample median values ($M_\mathrm{s,0}$, $r_\mathrm{s,0}$, and $T_\mathrm{0}$). We note that both $\alpha$ and $\beta$ are not affected by the choices of $T_0$, $M_\mathrm{s,0}$, and $r_\mathrm{s,0}$, only the normalization $\delta$ shifts accordingly.

One of the goals of this work is to determine the $\alpha$ and $\beta$ parameters that characterise the CFP (equation~(\ref{eq:plane})) in the most reliable way. To this end, we have explored several methods to perform fitting of the CFP, such as the $\chi^2$-minimisation method, the principal component analysis (PCA), the bidimensional least-square minimisation, and the minimum distance method (hereafter MINDISQ). The main difference among them is the choice of the parameter or reference quantity used to minimise and find the best-fit plane. We examined in detail the results obtained from the four fitting methods to conclude that the most reliable method for the present study is the MINDISQ method. For this method, the CFP corresponds to the plane that minimises the total sum of squared distances to the distribution of points, $\delta s^2:=\sum_i d_i^2$, where $|d_i|$ is the distance of each cluster to the CFP and $d_i$ is expressed as
\begin{equation}\label{eq:mindisq}
d_i = \frac{\alpha\,\log_{10}M_{i,\mathrm{s}} + \beta\,\log_{10}r_{i,\mathrm{s}} - \log_{10}T_i + \delta}{\sqrt{\alpha^2+\beta^2+1}}\ .
\end{equation}
By construction, this method provides the lowest dispersion or thickness ($\sigma_\mathrm{d}$) of the distribution of points around the resulting plane, which we define as half of the difference between the $84^\mathrm{th}$ and $16^\mathrm{th}$ percentiles of the distribution of $d_i$ values (distances to the best-fit CFP). We note that points above (below) the CFP yield positive (negative) $d_i$ values. 

Using the MINDISQ method, we compute the CFP of the simulated clusters from \three\ (Section~\ref{sec:fplane300}), as well as of the real clusters from CLASH (Section~\ref{sec:fplaneclash}) to compare these results. It is worth mentioning that the $\alpha$ and $\beta$ values obtained via the MINDISQ method are found to be quite similar to the results using PCA techniques. The main difference between the two methods is that PCA assumes that the plane is centred on the point $(M_\mathrm{s,0}$, $r_\mathrm{s,0}$,  $T_0$), which acts as a pivot point. This is not the case for the MINDISQ method because of the additional degree of freedom, $\delta$, which in turn avoids the possibility that $\alpha$ and $\beta$ are affected by the particular choice of $T_0$, $M_\mathrm{s,0}$, and $r_\mathrm{s,0}$. For this reason and to be consistent with the CFP coefficients obtained for the \three\ clusters, our CFP analysis of the CLASH sample is performed using the MINDISQ method in this work.

In addition, we explored the uncertainties of the CFP parameters, as well as their correlations. For the \three\ sample, we performed a bootstrapping analysis. In particular, we built $10^4$ bootstrap samples composed of $245$ clusters by using sampling with replacement from the simulated clusters with $\sigma_\mathrm{fit} \le 2$. For each of the bootstrap samples, we obtained a set of $\alpha$, $\beta$, and $\delta$ values by the MINDISQ method. The uncertainties and correlations of the parameters are obtained by the three sets of $10^4$ values. On the other hand, to estimate the uncertainties on the CFP parameters for the CLASH sample, we used the marginalised posterior distributions of \ms\ and \rs\ obtained by \citet[][$10^4$ steps]{Umetsu2016} and assumed uncorrelated Gaussian errors for the X-ray temperature obtained from \textit{Chandra} \citep[][]{Postman2012}.


\section{Fundamental plane in \three\ at $z=0$}\label{sec:fplane300}


We selected clusters at $z=0$ from \three\ according to the goodness of fit of the NFW model, $\sigma_\mathrm{fit} \le 2$, resulting in a sample composed of $245$ clusters. The sample spans characteristic halo masses of $M_\mathrm{s} \in [0.14, 7.05]\times 10^{14}h^{-1}~M_\odot$ and characteristic scale radii of $r_\mathrm{s}\in [149, 1003]\times h^{-1}$~kpc, whose median values are $M_\mathrm{s,0} = 2.03\times 10^{14}h^{-1}~M_\odot$ and $r_\mathrm{s,0}=424h^{-1}$~kpc, respectively. Depending on the definition of the weighted temperature, as well as the region in which this temperature is computed, the gas temperatures of the sample vary within the range $T\in [1.1, 15.2]$~keV (the minimum value refers to the spectroscopic-like temperature within $r_{200}$ and the maximum to the mass-weighted temperature within $r<500h^{-1}$~kpc). In this sense, the median gas temperature ranges from $5.0$~keV to $6.7$~keV and gas temperatures in inner regions ($r<500h^{-1}$~kpc) exhibit higher values than those within $r_{500}$ and $r_{200}$ (the temperature median values in Table~\ref{tab:plane300}). 

\begin{figure*}
    \centering
    \resizebox{\textwidth}{!}{\includegraphics[clip=True,trim=1.cm 0 0.cm 0]{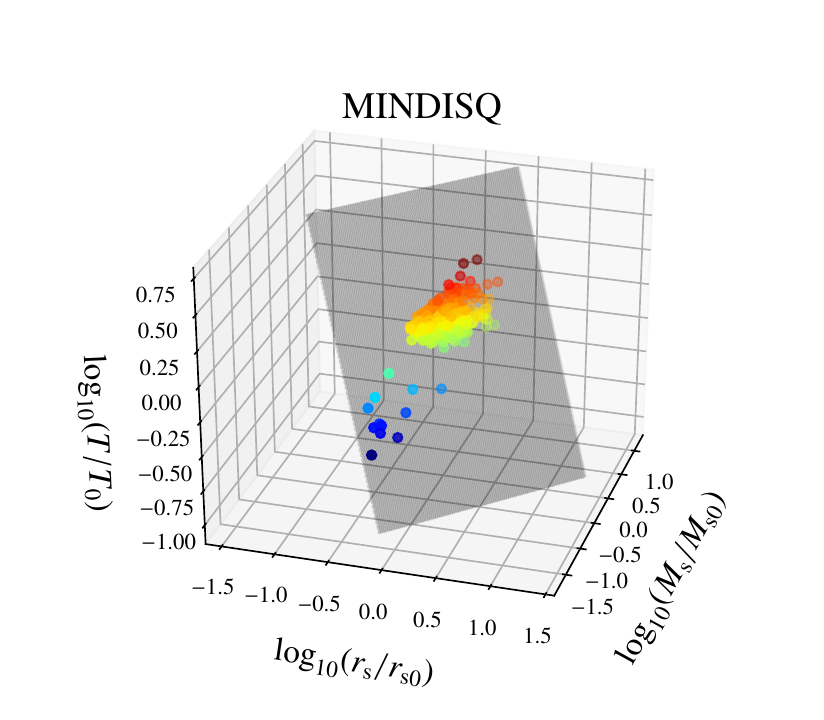}\includegraphics[clip=True,trim=1.cm 0 0.cm 0]{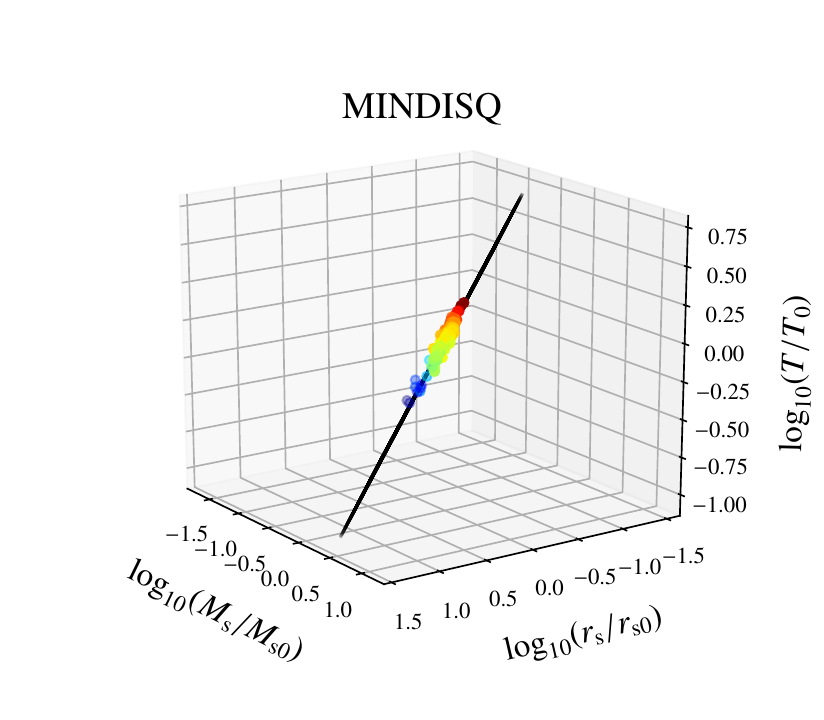}}
    \caption{Distribution of $245$ simulated galaxy clusters at $z=0$ from \three\ with $\sigma_\mathrm{fit}\le 2$ across the fundamental plane defined in the $(\log_{10}M_\mathrm{s}, \log_{10}r_\mathrm{s}, \log_{10}T)$ space where $M_\mathrm{s,0}=2.03\times10^{14}h^{-1}$~$M_\odot$, $r_\mathrm{s,0}=424h^{-1}$~kpc, and $T_0=5.05$~keV. All clusters are colour-coded according to their mass-weighted temperatures ($\log_{10}T/T_\mathrm{0}$) measured in the radial range [$0.1, 1.0$]$\times r_{200}$. \textit{Left-hand side panel}: best-fit plane (grey) obtained with the MINDISQ method (further details in the text). \textit{Right-hand side panel}: same as \textit{left-hand side panel}, but from a different viewing angle to illustrate the low dispersion of the clusters with respect to the best-fit plane.
    \label{fig:fplane_300}}
\end{figure*}

In Table~\ref{tab:plane300}, we illustrate the results for all gas temperature definitions for \three\ obtained with the MINDISQ method. The first conclusion from the MINDISQ analysis (also obtained for the other methods explored in this work) is that the clusters from \three\ lie on a thin plane, as shown in Fig.~\ref{fig:fplane_300}. Secondly, the dispersion or thickness of this plane depends on the gas temperature definition and it spans a range of values of $\sigma_\mathrm{d}\in [0.015, 0.030]$~dex (average thickness of $0.02$~dex; see Fig.~\ref{fig:fplane_300} and Table~\ref{tab:plane300}). The thickness is systematically lower for the mass-weighted gas temperature than for the spectroscopic-like one. Thirdly, the values of $\alpha$, $\beta$, and $\delta$ that define the CFP (see equation~(\ref{eq:plane})) also vary according to the gas temperature definition employed. 

Regarding the best-fit values of $\alpha$ and $\beta$, it should be noted that for all temperature definitions in \three,  we find $1 < \alpha < 1.5$ and $-2 < \beta < -1$, that is, halfway between the expectation for simplified virial equilibrium and the similarity solution for a secondary infall model. In fact, the case for the mass-weighted temperature within $r_{200}$ ($T_\mathrm{mw}(r_{200})$, see Table~\ref{tab:plane300}) is fully compatible with the simplified virial expectation within a $1\sigma$ uncertainty level. In general, we find that the \three\ results for the mass-weighted temperature are closer to the virial expectation, while the results for the spectroscopic-like temperature show intermediate values between the virial expectation and the similarity solution. 

\begin{figure}
    \centering
    \resizebox{\columnwidth}{!}{\includegraphics[clip=True,trim=0.4cm 0 1cm 0]{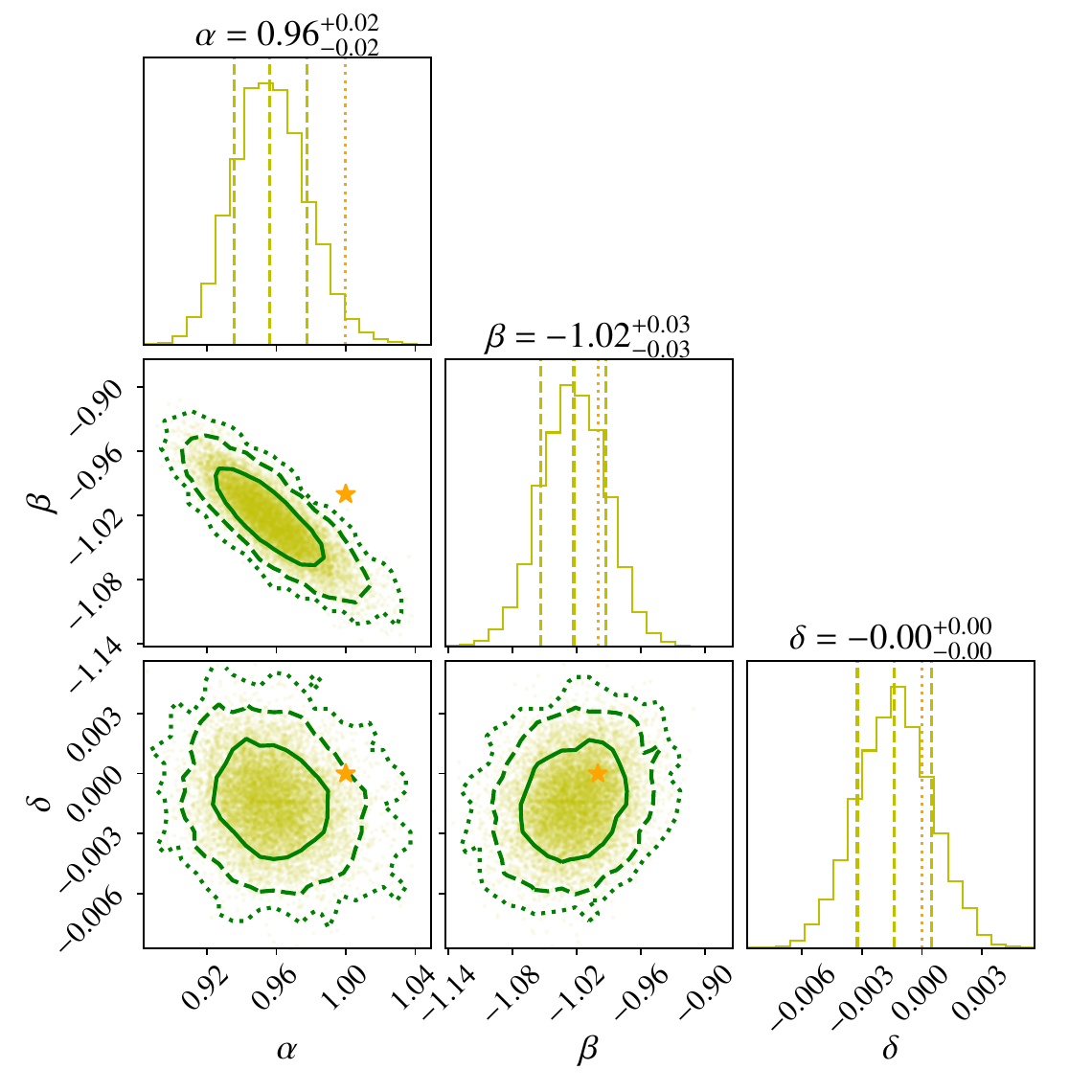}}
    \caption{Constraints on the parameters ($\alpha,\beta,\delta$) defining the fundamental plane (see equation~(\ref{eq:plane})) of galaxy clusters at $z=0$ from \three\ obtained using mass-weighted temperatures within $r_{200}$. Green solid, dashed, and dotted lines illustrate the confidence levels of $68$, $95$, and $99.7$ per cent of probability, respectively. Dashed yellow lines denote the $16^\mathrm{th}$, $50^\mathrm{th}$, and $84^\mathrm{th}$ percentiles of the distributions of parameters obtained from bootstrapping. In each panel, the orange star-shaped symbol or the vertical dotted line indicates the expectation for the simplified virial equilibrium case.
    \label{fig:mc300}}
\end{figure}

There is evidence of systematic trends between ($\alpha$, $\beta$) and the temperature definition used to define the CFP (Table~\ref{tab:plane300}). First, the $\alpha$ and $\beta$ values are similar in the cases of $T(r_{500})$ and $T(r_{200})$, in which the temperatures were measured within apertures that are much larger than the halo scale radius, $r_\mathrm{s}$. For $T(500h^{-1}~\mathrm{kpc})$ where gas temperatures were measured within a fixed aperture that is close to the median value of $r_\mathrm{s}$, the resulting CFP deviates the most from the virial expectation and shifts towards the similarity solution. Second, the CFPs obtained with $T_\mathrm{sl}$ exhibit systematically higher (lower) values of $\alpha$ ($\beta$) compared to those obtained with $T_\mathrm{mw}$ computed within the same radial range. As a result, the CFPs with $T_\mathrm{sl}$ tend to be more shifted towards the similarity solution than those with $T_\mathrm{mw}$. Since the spectroscopic-like temperature is more sensitive to the inner ICM region than the mass-weighted one, both trends indicate that the systematic deviations from the virial expectation are likely caused by the inner temperature distribution ($r\lesssim r_\mathrm{s}$). We note that $\alpha$ and $\beta$ are anticorrelated, whereas $\delta$ appears to be uncorrelated with respect to the other two parameters (Fig.~\ref{fig:mc300}).

We also derive a CFP for clusters with spectroscopic-like temperatures of $T_\mathrm{sl}(500h^{-1}~\mathrm{kpc}) > 5~\mathrm{keV}$, mimicking the CLASH X-ray selection (see Table~\ref{tab:plane300} and Fig.~\ref{fig:comparison_sample}). There are $201$ clusters in \three\ with $T_\mathrm{sl}(500h^{-1}~\mathrm{kpc}) > 5~\mathrm{keV}$ and $\sigma_\mathrm{fit} \le 2$. We find that the $\alpha$ and $\beta$ values mildly change with respect to the ones obtained for the full sample in the same region, reaching a similar conclusion.

\begin{table*}
\centering
\caption{Best-fit parameters ($\alpha,\beta,\delta$) and thickness ($\sigma_\mathrm{d}$ in dex units) of the fundamental plane defined in the $(\log_{10}r_\mathrm{s}, \log_{10}M_\mathrm{s}, \log_{10}T)$ space obtained for a sample of $245$ simulated galaxy clusters at $z=0$ (see equation~(\ref{eq:plane})) from \three\ with $\sigma_\mathrm{fit}\le 2$. The results are shown separately for the mass-weighted and spectroscopic-like temperatures ($T_\mathrm{mw}$ and $T_\mathrm{sl}$, respectively) computed within a radial range of $500h^{-1}$~kpc, $r_{500}$, and $r_{200}$ (see further details in the text). All results were obtained by setting $M_\mathrm{s,0}=2.03\times 10^{14}h^{-1}~M_\odot$ and $r_\mathrm{s,0} = 424h^{-1}$~kpc. For each quantity, the lower and upper errors enclose the $1\sigma$ uncertainty range.
\label{tab:plane300}}
    
\begin{tabular}{lccccc}
\hline
 & $\alpha$ & $\beta$ & $\delta$ & $\sigma_\mathrm{d}$ & $T_\mathrm{0}$\\
 & & & & [dex] & [keV] \\
\hline
$T_\mathrm{mw}(500h^{-1}~\mathrm{kpc})$ & $1.10^{+0.04}_{-0.03}$ & $-1.22^{+0.05}_{-0.05}$ & $\ \ 0.001^{+0.003}_{-0.003}$ & $0.021$ & $6.66$\\
$T_\mathrm{mw}(r_{500})$         & $0.99^{+0.02}_{-0.02}$ & $-1.09^{+0.03}_{-0.03}$ & $-0.001^{+0.002}_{-0.002}$ & $0.015$ & $5.62$\\
$T_\mathrm{mw}(r_{200})$         & $0.96^{+0.02}_{-0.02}$ & $-1.02^{+0.03}_{-0.03}$ & $-0.001^{+0.002}_{-0.002}$ & $0.015$ & $5.05$\\
$T_\mathrm{sl}(500h^{-1}~\mathrm{kpc})$ & $1.22^{+0.06}_{-0.05}$ & $-1.50^{+0.08}_{-0.09}$ & $-0.007^{+0.004}_{-0.004}$ & $0.030$ & $6.26$\\
$T_\mathrm{sl}(r_{500})$           & $1.10^{+0.04}_{-0.03}$ & $-1.36^{+0.05}_{-0.05}$ & $-0.003^{+0.003}_{-0.003}$ & $0.018$ & $5.54$\\
$T_\mathrm{sl}(r_{200})$           & $1.12^{+0.04}_{-0.03}$ & $-1.43^{+0.04}_{-0.05}$ & $-0.006^{+0.003}_{-0.003}$ & $0.020$ & $5.32$\\
$T_\mathrm{sl}(500h^{-1}~\mathrm{kpc})> 5~\mathrm{keV}$           & $1.27^{+0.07}_{-0.06}$ & $-1.45^{+0.07}_{-0.08}$ & $-0.012^{+0.004}_{-0.004}$ & $0.026$ & $6.63$\\
\hline
\multicolumn{2}{c}{Number=$245$}  & \multicolumn{2}{c}{$M_\mathrm{s,0}=2.03 \times 10^{14}h^{-1}~M_\odot$} & \multicolumn{2}{c}{$r_\mathrm{s,0}=424h^{-1}$~kpc} \\
\hline
\end{tabular}

\end{table*}

\section{The fundamental plane of clusters at different redshift}\label{sec:cfp_z}


The redshift evolution of the CFP is explored using clusters from the \three. For this aim, we use our set of characteristic parameters and gas temperatures for the simulated clusters at higher redshifts in the simulation ($z=0.07, 0.22, 0.33, 0.59$, and $0.99$). As in previous sections, the CFP at each redshift is determined using the MINDISQ method and only using those clusters that were properly fitted by an NFW profile (i.e., $\sigma_\mathrm{fit} \le 2$). As a result, the number of clusters used to determine the CFP at each redshift can vary: at $z=0.99$, there are $280$ clusters, then the number of objects decreases down to $231$ for $z=0.07$.  For each redshift, the uncertainties and correlations of the $\alpha$,  $\beta$, and $\delta$ parameters were determined using $10^4$ bootstrap samples.

As summarised in Table~\ref{tab:cfp_allz} (see Appendix~\ref{sec:tables}), the best-fit $\alpha$ and $\beta$ parameters defining the CFP span the ranges of $0.9 < \alpha < 1.3$ and $-1.6 < \beta < -1$ at $z<1$.  As found for the CFP at $z=0$ (Section~\ref{sec:fplane300}),  the best-fit values of $\alpha$ and $\beta$ depend on the gas temperature definition (mass-weighted and spectroscopic-like) and on their radial ranges (see Table~\ref{tab:cfp_allz} in Appendix~\ref{sec:tables}). The absolute values of $\alpha$ and $\beta$ are slightly larger for the spectroscopic-like temperature with respect to the mass-weighted one. This is especially true for the outer regions ($r_{500}$ and $r_{200}$) compared to that obtained for the radial range of $500h^{-1}$~kpc. We notice that for the mass-weighted temperature inside $r_{200}$, $T_\mathrm{mw}(r_{200})$, the CFP is very close to the virial expectation ($\alpha=1$ and $\beta=-1$). However, we stress again that $T_\mathrm{mw}(r_{200})$ is measured within an aperture that is much larger than the halo scale radius $r_\mathrm{s}$ (see the distribution of $c_{200}=r_{200}/r_\mathrm{s}$ in the right panel of Fig.~\ref{fig:comparison_sample}).

Regarding the CFP thickness, we find that there are subtle differences between the mass-weighted and the spectroscopic-like temperatures, where $\sigma_\mathrm{d}$ tends to be marginally higher (of the order of $\sim 10$~per-cent) for the latter.  


To quantify the evolution of the CFP with redshift, we assume that $\alpha$ and $\beta$ are redshift-dependent parameters.  Specifically, we parametrise the redshift evolution of the CFP parameters as follows: $\alpha(z) = \alpha_0 \ (1+z)^{\alpha_1}$  and $\beta(z) = \beta_0 \ (1+z)^{\beta_1}$. As a consequence, $\alpha_1 = \beta_1 = 0$ implies no evolution of the CFP.  To determine the uncertainties of the parameters, we performed this fit on the bootstrap samples, obtaining $10^4$  sets of ($\alpha_0$, $\alpha_1$, $\beta_0$, $\beta_1$) values for each of the temperatures and radial ranges (see Table~\ref{tab:coeff_z_log}).

\begin{table}
\centering
\caption{Redshfit-dependent coefficients of the fundamental plane, $\alpha(z)=\alpha_0\,(1+z)^{\alpha_1}$ and $\beta(z)=\beta_0\,(1+z)^{\beta_1}$, for \three\ galaxy clusters.
\label{tab:coeff_z_log}}

\begin{tabular}{lcccc}
\hline
 & $\alpha_1$ & $\alpha_0$ & $\beta_1$ & $\beta_0$ \\
\hline
$T_\mathrm{mw}(500h^{-1}~\mathrm{kpc})$ & $-0.11^{+0.04}_{-0.04}$ & $1.15^{+0.02}_{-0.02}$ & $-0.14^{+0.06}_{-0.05}$ & $-1.30^{+0.03}_{-0.03}$ \\
$T_\mathrm{mw}(r_{500})$ & $-0.17^{+0.04}_{-0.04}$ & $1.05^{+0.02}_{-0.02}$ & $-0.19^{+0.05}_{-0.05}$ & $-1.19^{+0.03}_{-0.03}$ \\
$T_\mathrm{mw}(r_{200})$ & $-0.13^{+0.04}_{-0.04}$ & $1.01^{+0.02}_{-0.02}$ & $-0.16^{+0.06}_{-0.06}$ & $-1.12^{+0.03}_{-0.03}$ \\
$T_\mathrm{sl}(500h^{-1}~\mathrm{kpc})$ & $-0.26^{+0.05}_{-0.06}$ & $1.22^{+0.03}_{-0.03}$ & $-0.24^{+0.07}_{-0.07}$ & $-1.47^{+0.04}_{-0.05}$ \\
$T_\mathrm{sl}(r_{500})$ & $-0.34^{+0.05}_{-0.05}$ & $1.19^{+0.03}_{-0.03}$ & $-0.39^{+0.06}_{-0.07}$ & $-1.50^{+0.04}_{-0.04}$ \\
$T_\mathrm{sl}(r_{200})$ & $-0.35^{+0.05}_{-0.05}$ & $1.23^{+0.03}_{-0.02}$ & $-0.40^{+0.06}_{-0.06}$ & $-1.56^{+0.04}_{-0.04}$ \\
\hline
\end{tabular}

\end{table}

\begin{figure}
\centering
    \includegraphics[width=0.8\columnwidth,clip=True]{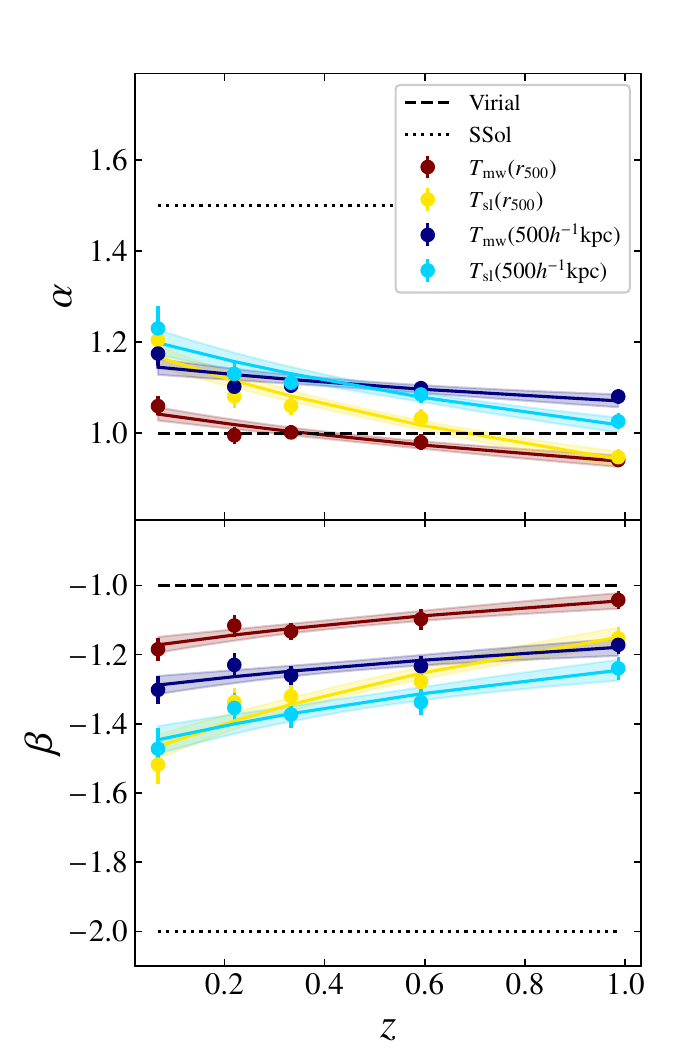}
    \caption{Evolution of the parameters $\alpha$ and $\beta$ defining the fundamental plane for \three\ galaxy clusters as a function of redshift. The results are shown for different temperature definitions and radial ranges (see inset). The dashed and dotted black lines indicate the virial expectation and the similarity solution for a secondary infall model, respectively.
    \label{fig:coeff_z}}
\end{figure}

According to \three\ simulations, there are hints for an evolution of $\alpha$ and $\beta$ since $z\sim1$ (see Table~\ref{tab:coeff_z_log} and Fig.~\ref{fig:coeff_z}).  Specifically, our results point out that the CFP may gradually change with redshift and this trend is consistent for all temperature definitions and radial ranges probed. Using the mass-weighted temperature, the evolution of the two parameters, $\alpha$ and $\beta$, is mild with a maximum of $15$~per cent between $z=1$ and $z=0$, while in the same temporal range they vary with a maximum of $25$~per cent for the spectroscopic-like temperature. It is noteworthy that based on \three\ simulations, we would expect that clusters form a CFP similar to the virial expectation at $z=1$, which evolves into a CFP that is halfway between the virial expectation and the similarity solution for a secondary infall model. In general, all cases exhibit $\alpha$ and $\beta$ closer to the virial expectation ($\alpha=1$ and $\beta=-1$) at higher redshifts. Notice that at all times, the scatter around the CFP remains very low and constant, implying that the CFP remains well defined throughout the evolution of the cluster population.

\section{Dynamical relaxation}\label{sec:relaxation}


During the growth history of individual halos, the structural parameters of the NFW profile can be affected by the accretion of matter from their surroundings.  The characteristic parameters \rs\ and \ms\ of individual clusters may thus change during its assembly. Then, the CFP may well be different between relaxed and unrelaxed clusters. Here we investigate whether the CFP depends on the dynamical relaxation state of galaxy clusters (see Sections~\ref{sec:fplane300} and \ref{sec:cfp_z}) using \three\ clusters.

For this purpose, we use the relaxation parameters defined in \citet{Cui2018}: the virial ratio $\eta$, the centre-of-mass offset $\Delta_\mathrm{r}$, and the fraction of mass $f_\mathrm{s}$ (see Section~\ref{sec:method_dynamical} for their definitions).  Following \citet{Cui2018}, we assume that a cluster is dynamically relaxed  when it satisfies that $0.85 < \eta < 1.15$, $\Delta_\mathrm{r} < 0.4$, and $f_\mathrm{s} < 0.1$. We note that these limiting values are defined inside $r_{200}$. To facilitate the simultaneous use of these three indicators, we combine them to define a new indicator or relaxation coefficient \relaxtext\ (see equation~(\ref{eq:relax})), which is higher than unity for relaxed clusters. 

\begin{figure}
\centering
    \resizebox{\columnwidth}{!}{\includegraphics[clip=True]{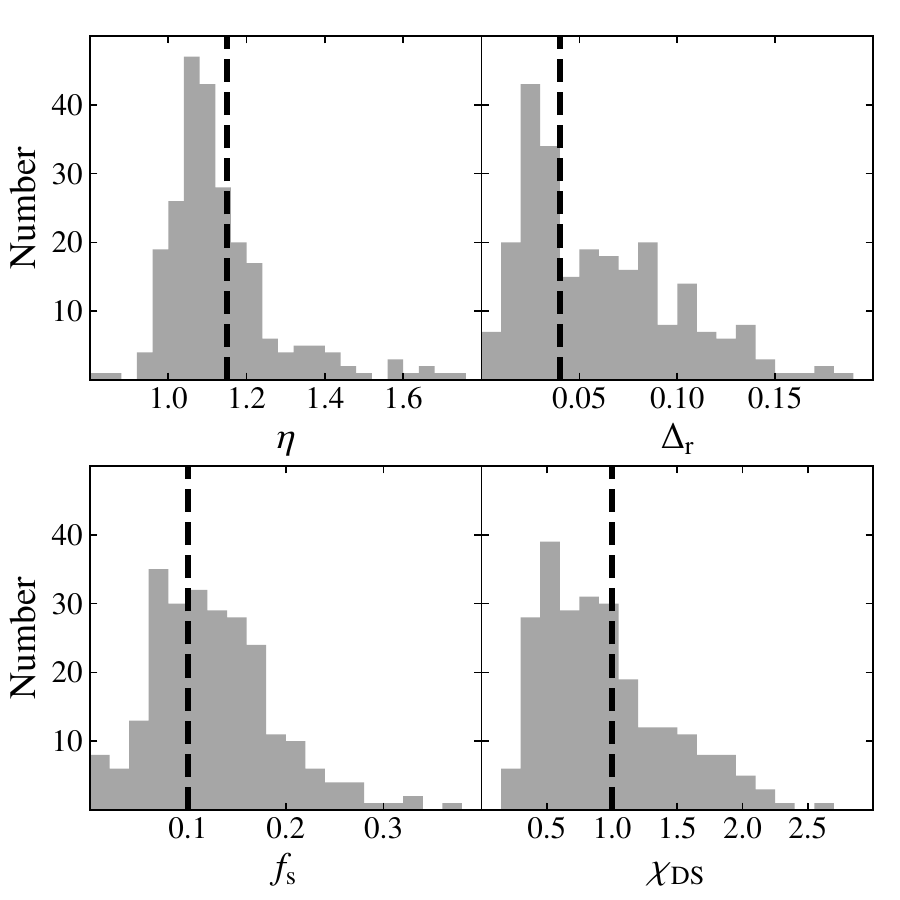}}
    \caption{
    Distributions of the four dynamical relaxation indicators of \three\ clusters at $z=0$. \textit{Upper panels}: virial ratio ($\eta$) and centre-of-mass offset ($\Delta_\mathrm{r}$). \textit{Lower panels}: fraction of mass in subhalos ($f_\mathrm{s}$) and the combination of the three parameters (\relaxtext, see equation~(\ref{eq:relax})). The dashed lines show the threshold values for classification as `relaxed' objects.
    \label{fig:relax_hist}}
\end{figure}

\begin{figure*}
\centering
    \resizebox{\textwidth}{!}{\includegraphics[clip=True]{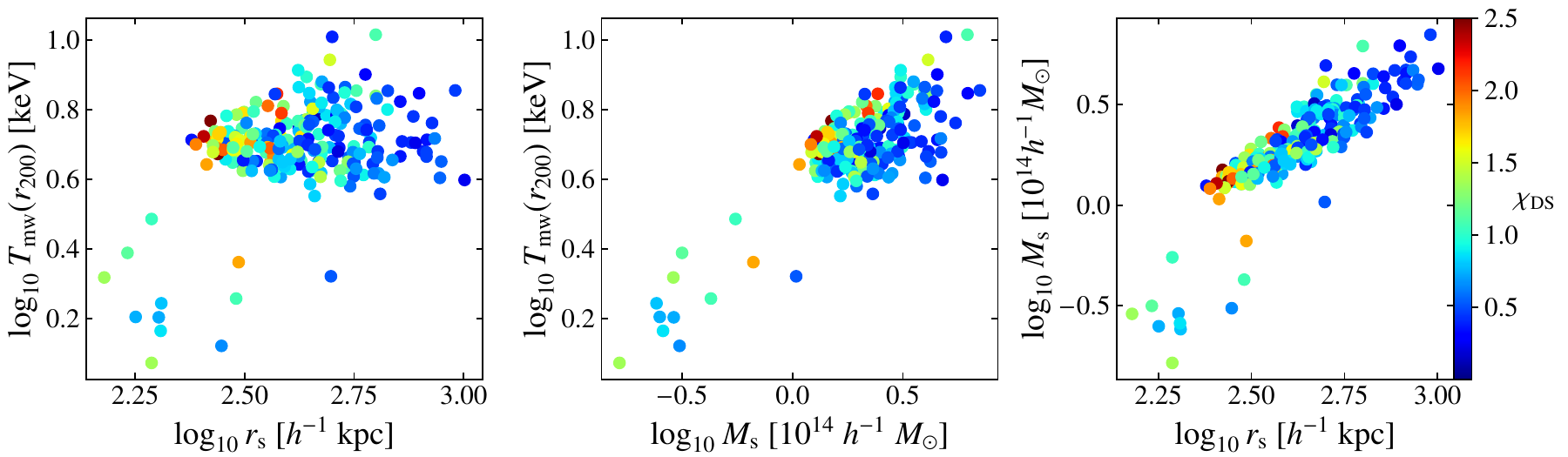}}
    \caption{
    Distribution of the dynamical relaxation parameter \relaxtext\ (see equation~(\ref{eq:relax})) in three projections of the ($\log_{10}M_\mathrm{s}, \log_{10}r_\mathrm{s}, \log_{10}T$) space for \three\ clusters at $z=0$. Redder (bluer) colours correspond to relaxed (unrelaxed) clusters.
    \label{fig:relax_cfp}}
\end{figure*}

We find that, overall, $37$ per-cent of the sample is dynamically relaxed ($\relaxmath \ge 1$; see the bottom-right panel of Fig.~\ref{fig:relax_hist}).  Although the fraction of relaxed and unrelaxed clusters changes according to the relaxation criteria employed and on the halo mass as shown in \citet[][]{Cui2018},  the sample contains more unrelaxed clusters than relaxed ones. The fraction of relaxed clusters increases with decreasing halo mass \citep[see also Table~4 in][]{Cui2018}. 

We find that relaxed and unrelaxed clusters at $z=0$ (see Fig.~\ref{fig:relax_cfp}) lie in well defined regions of the CFP. Unrelaxed clusters exhibit, on average, higher values of $r_\mathrm{s}$ and $M_\mathrm{s}$, whereas relaxed clusters typically have lower values of $r_\mathrm{s}$ and $M_\mathrm{s}$, populating the opposite `side' of the CFP. However, we do not find that \three\ clusters preferentially lie in a certain range of gas temperatures according to the degree of dynamical relaxation. These indicate that the CFP depends on the degree of dynamical relaxation of clusters, which should be explored in more detail.  To this end, we split our sample of clusters into two subsamples according to the relaxation coefficient \relaxtext. Hereafter, those clusters with $\relaxmath \ge 1$ are referred as relaxed systems, while those with $\relaxmath \le 2/3$ are referred as unrelaxed systems.


\subsection{The CFP of relaxed and unrelaxed clusters at $z=0$}\label{sec:relax_z0}

In our sample of $245$ clusters at $z=0$, there are $90$ clusters with $\relaxmath \ge 1$.  We repeat the analysis of CFP fitting using only relaxed clusters that are properly fitted by an NFW profile (i.e., $\sigma_\mathrm{fit} \le 2$). The results are summarised in Table~\ref{tab:plane300_relax} in Appendix~\ref{sec:tables}. We find that the CFP of relaxed clusters is closer to the expectation of simplified virial equilibrium ($\alpha=1$ and $\beta=-1$), compared to the CFP obtained for the full sample (Section~\ref{sec:fplane300}). This result is independent of the definition of weighted gas temperature and the radial ranged employed. In particular, the CFP parameters obtained using the mass-weighted temperature are compatible with $\alpha=1$ and $\beta=-1$, even for the radial range of $500h^{-1}$~kpc. For the spectroscopic-like temperature, we also find the same systematic differences with respect to the results obtained for the full sample  (Section~\ref{sec:fplane300}). The $\alpha$ parameter is typically smaller by $\sim 0.06 \pm 0.03$ than for the full sample, while the $\beta$ parameter is larger (less negative) by $\sim 0.15 \pm 0.04$ than for the full sample (see Tables~\ref{tab:plane300} and \ref{tab:plane300_relax}). The thickness of the CFP of relaxed clusters is also systematically smaller than obtained for the full sample by down to $0.005$~dex (see Table~\ref{tab:plane300_relax} in Appendix~\ref{sec:tables}).


Regarding the unrelaxed subsample,  there are $87$ clusters with $\relaxmath \le 2/3$ at $z=0$. The trends in the changes in $\alpha$ and $\beta$ are opposite to those for the relaxed subsample (see Table~\ref{tab:plane300_relax} in Appendix~\ref{sec:tables}). Overall, except for the mass-weighted temperatures within $r_{500}$ and $r_{200}$, $\alpha$ slightly increases by $0.10 \pm 0.05$, while $\beta$ decreases by $0.15 \pm 0.05$. It is remarkable that for the mass-weighted temperatures within $r_{500}$ and $r_{200}$, $\alpha$ and $\beta$ are close to virial expectations.

We find that $\beta$ is the CFP parameter most sensitive to the relaxation classification (see Table~\ref{tab:plane300_relax} in Appendix~\ref{sec:tables}). The CFP of unrelaxed clusters is halfway between the virial expectation and the similarity solution. It should be noted that for the spectroscopic-like temperature with the $500h^{-1}$~kpc radial range, the $\alpha$ and $\beta$ values are close to the similarity solution ($\alpha=1.5$ and $\beta=-2$). As expected, the thickness of the CFP for the unrelaxed subsample is slightly larger than obtained for the full sample and the relaxed subsample. Hence, although clusters systematically move across the $(\log{M_\mathrm{s}}, \log{r_\mathrm{s}}, \log{T})$ space during their evolution, the mass growth of clusters can slightly tilt the CFP and increase the dispersion of this plane.


\subsection{Redshift evolution of the CFP for relaxed and unrelaxed clusters}\label{sec:relax_z} 

Here we explore the evolution of the CFP with redshift for relaxed and unrelaxed clusters separately. To this end, we use the \three\ clusters identified at different epochs  (Section~\ref{sec:cfp_z}) and at each redshift we split them into relaxed and unrelaxed subsamples according to our relaxation criteria  ($\relaxmath \ge 1$ and $\relaxmath \le 2/3$, respectively).  Specifically, we derive the CFP for each subsample at $z=0.07$, $0.22$, $0.33$, $0.59$, and $0.99$.  We note that the relaxation parameter \relaxtext\ for each individual cluster is recomputed within $r_{200}$ at each redshift, so that some clusters identified as relaxed at $z=0$ may be identified as unrelaxed at higher redshifts, and vice versa. As in Section~\ref{sec:cfp_z}, the redshift evolution of the CFP parameters $\alpha(z)$ and $\beta(z)$ is parametrised as $\alpha(z) = \alpha_0 \ (1+z)^{\alpha_1}$  and $\beta(z) = \beta_0 \ (1+z)^{\beta_1}$.

\begin{figure*}
\centering
\includegraphics[width=0.8\columnwidth,clip=True]{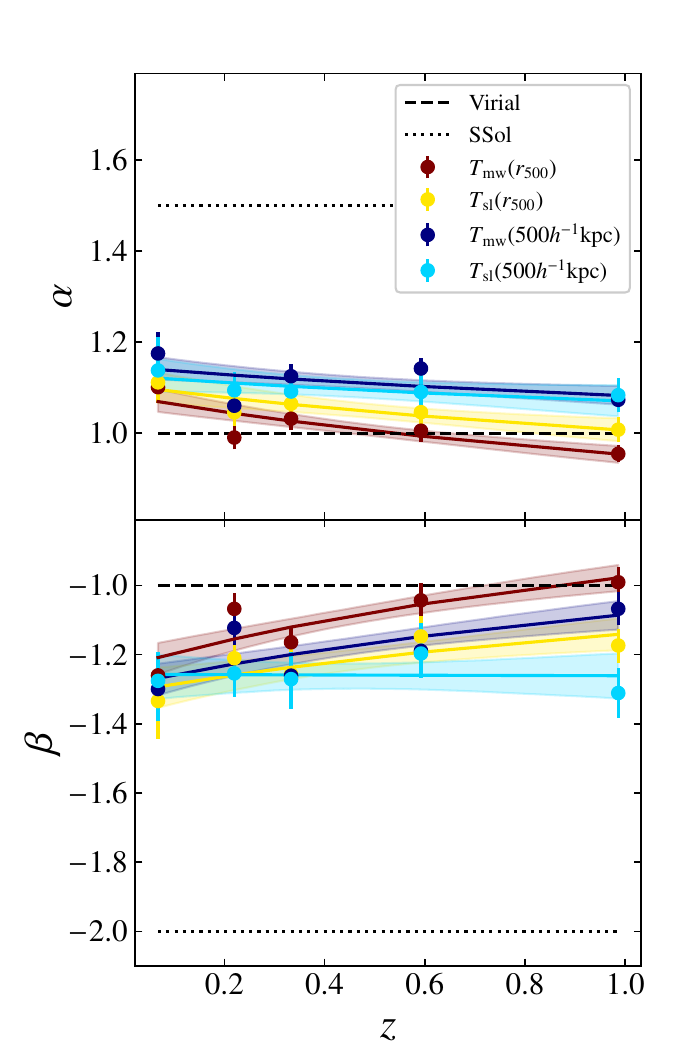}
\includegraphics[width=0.8\columnwidth,clip=True]{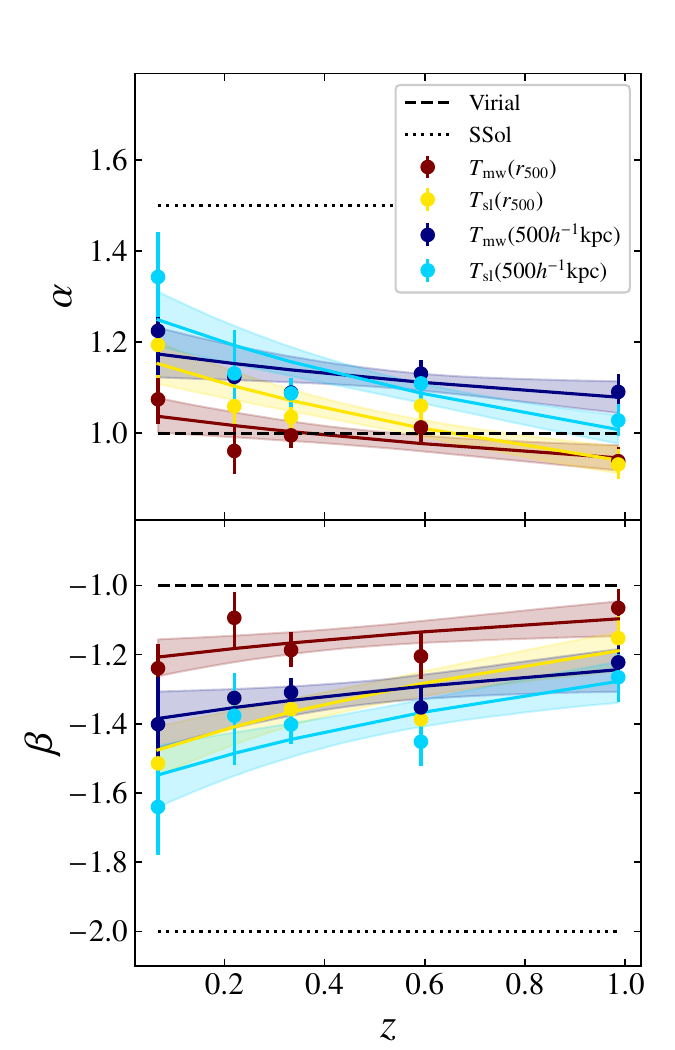}
\caption{Same as Fig.~\ref{fig:coeff_z}, but for relaxed and unrelaxed clusters (left and right panels, respectively).
\label{fig:coeff_z_relax}}
\end{figure*}

There are more (fewer) relaxed (unrelaxed) clusters at $z=0$ than at $z=1$, all selected to have $\sigma_\mathrm{fit} \le 2$.  We find that the redshift evolution of $\alpha$ and $\beta$ for the relaxed subsample is milder than for the unrelaxed subsample (see Fig.~\ref{fig:coeff_z_relax} and Table~\ref{tab:coeff_z_log_relax}). Moreover, the CFP parameters $\alpha$ and $\beta$ for the relaxed subsample are closer to the virial expectation at all redshifts probed. On the other hand, the unrelaxed subsample exhibits a stronger evolution for the spectroscopic-like temperature case. The CFP at $z\sim 1$ is close to the virial expectation as found in Section~\ref{sec:cfp_z}. However, the absolute values of $\alpha$ and $\beta$ for the unrelaxed subsample increase progressively towards lower redshifts (see the right panel of Fig.~\ref{fig:coeff_z_relax}), coming closer to the similarity solution for a secondary infall model. Our results thus suggest that relaxed and unrelaxed clusters qualitatively evolve in a similar manner (growing $\alpha$ and decreasing $\beta$ towards $z=0$), but with different amplitude.

\begin{table}
\centering
\caption{Same as Table~\ref{tab:coeff_z_log}, but for the relaxed ($\relaxmath \ge 1$, upper panel) and unrelaxed ($\relaxmath \le 2/3$, lower panel) subsamples.
\label{tab:coeff_z_log_relax}}
    
\begin{tabular}{l cccc }
\hline
 $\relaxmath \ge 1$ & $\alpha_1$ & $\alpha_0$ & $\beta_1$ & $\beta_0$ \\
\hline
$T_\mathrm{mw}(500h^{-1}~\mathrm{kpc})$ & $-0.08^{+0.05}_{-0.06}$ & $1.15^{+0.03}_{-0.02}$ & $-0.25^{+0.10}_{-0.10}$ & $-1.29^{+0.05}_{-0.06}$  \\
$T_\mathrm{mw}(r_{500})$ & $-0.18^{+0.05}_{-0.06}$ & $1.08^{+0.03}_{-0.03}$ & $-0.34^{+0.10}_{-0.11}$ & $-1.24^{+0.05}_{-0.06}$  \\
$T_\mathrm{mw}(r_{200})$ & $-0.20^{+0.05}_{-0.05}$ & $1.05^{+0.03}_{-0.02}$ & $-0.41^{+0.10}_{-0.10}$ & $-1.18^{+0.05}_{-0.05}$  \\
$T_\mathrm{sl}(500h^{-1}~\mathrm{kpc})$ & $-0.07^{+0.08}_{-0.09}$ & $1.13^{+0.05}_{-0.03}$ & $\ \ 0.01^{+0.13}_{-0.14}$ & $-1.26^{+0.07}_{-0.08}$  \\
$T_\mathrm{sl}(r_{500})$ & $-0.14^{+0.07}_{-0.08}$ & $1.11^{+0.04}_{-0.03}$ & $-0.20^{+0.11}_{-0.12}$ & $-1.31^{+0.06}_{-0.07}$  \\
$T_\mathrm{sl}(r_{200})$ & $-0.20^{+0.07}_{-0.08}$ & $1.14^{+0.05}_{-0.03}$ & $-0.33^{+0.11}_{-0.12}$ & $-1.40^{+0.07}_{-0.08}$  \\
\hline
\end{tabular}

\begin{tabular}{l cccc }
$\relaxmath \le 2/3$ & $\alpha_1$ & $\alpha_0$ & $\beta_1$ & $\beta_0$ \\
\hline
$T_\mathrm{mw}(500h^{-1}~\mathrm{kpc})$ & $-0.14^{+0.11}_{-0.12}$ & $1.18^{+0.07}_{-0.06}$ & $-0.17^{+0.15}_{-0.15}$ & $-1.40^{+0.09}_{-0.10}$ \\
$T_\mathrm{mw}(r_{500})$ & $-0.15^{+0.09}_{-0.09}$ & $1.05^{+0.05}_{-0.04}$ & $-0.16^{+0.12}_{-0.12}$ & $-1.22^{+0.06}_{-0.07}$ \\
$T_\mathrm{mw}(r_{200})$ & $-0.03^{+0.11}_{-0.11}$ & $0.98^{+0.05}_{-0.05}$ & $-0.06^{+0.16}_{-0.16}$ & $-1.11^{+0.08}_{-0.08}$ \\
$T_\mathrm{sl}(500h^{-1}~\mathrm{kpc})$ & $-0.35^{+0.11}_{-0.11}$ & $1.28^{+0.07}_{-0.07}$ & $-0.31^{+0.14}_{-0.14}$ & $-1.58^{+0.10}_{-0.11}$ \\
$T_\mathrm{sl}(r_{500})$ & $-0.33^{+0.10}_{-0.10}$ & $1.18^{+0.06}_{-0.05}$ & $-0.35^{+0.13}_{-0.13}$ & $-1.51^{+0.08}_{-0.09}$  \\
$T_\mathrm{sl}(r_{200})$ & $-0.40^{+0.10}_{-0.10}$ & $1.26^{+0.07}_{-0.06}$ & $-0.40^{+0.12}_{-0.13}$ & $-1.59^{+0.09}_{-0.10}$  \\
\hline
\end{tabular}

\end{table}


\subsection{Dependence of the CFP on the formation redshift}\label{sec:zform}

The formation redshift of clusters $z_\mathrm{form}$, defined as the time when the mass reaches half of the value at $z=0$ ($M_{200}(z_\mathrm{form})= 0.5 M_{200}(z=0)$), is linked to their mass assembly history and dynamical relaxation state. Clusters formed earlier have had more time to relax and have likely already transitioned from the fast to slow accretion phase, when the halo growth mostly occurs in their outskirts and \rs\ remains approximately constant. \citet{Mostoghiu2019} found that relaxed and unrelaxed clusters at $z=0$ from \three\ have different distributions of $z_\mathrm{form}$.  On average, relaxed clusters at $z=0$ have higher $z_\mathrm{form}$ compared to unrelaxed clusters. Using \three\ clusters at $z=0$ selected to have $\sigma_\mathrm{fit} \le 2$, we explore the CFP in three bins of formation redshift:  $z_\mathrm{form} < 0.4$, $0.4 \le z_\mathrm{form} < 0.6$, and $z_\mathrm{form} \ge 0.6$. The results are summarised in Table~\ref{tab:plane300_zform} in Appendix~\ref{sec:tables}.

Clusters that formed earlier, $z_\mathrm{form}\ge 0.6$, and thus those that are likely in the slow accretion phase, form a CFP at $z=0$ that is deviated from the similarity solution for a secondary infall model (see Table~\ref{tab:plane300_zform} in Appendix~\ref{sec:tables}). This conclusion applies to all temperature definitions and radial ranges explored in this work. In particular, the results obtained for the $z_\mathrm{form}\ge 0.6$ subsample using the mass-weighted temperature are fairly consistent with the expectation for simplified virial equilibrium ($\alpha=1$ and $\beta=-1$). It should be noted that we find qualitatively similar results for relaxed clusters (see Table~\ref{tab:plane300_relax} in Appendix~\ref{sec:tables}). This is not surprising, since clusters classified as relaxed ($\relaxmath \ge 1$) exhibit typical formation redshifts of $z_\mathrm{form} > 0.4$, although there are some unrelaxed clusters ($\relaxmath \le 2/3$) with $z_\mathrm{form} \ge 0.6$ and some relaxed clusters ($\relaxmath\ge 1$) with $z_\mathrm{form}<0.6$.  The CFP parameters obtained with the spectroscopic-like temperature are slightly closer to the similarity solution compared to those with the mass-weighted temperature.

On the other hand, the CFP parameters obtained for recently formed clusters with $z_\mathrm{form} < 0.4$ ($\alpha\sim 1.25$ and $\beta\sim -1.5$; see Table~\ref{tab:plane300_zform}) lie halfway between the simplified virial expectation and the similarity solution. In particular, the results obtained with $T_\mathrm{sl}(500h^{-1}\mathrm{kpc})$ are compatible with the similarity solution. These findings are in line with the results for the subsample of unrelaxed clusters ($\relaxmath \le 2/3$). This is a natural consequence of the fact that all \three\ clusters with $z_\mathrm{form} < 0.4$ have $\relaxmath < 1$ (see also Table~\ref{tab:plane300_relax} in Appendix~\ref{sec:tables}). 

For the subsample of intermediate-$z_\mathrm{form}$ clusters with $0.4 \le z_\mathrm{form} < 0.6$, we find CFP parameters that are consistent with the $z_\mathrm{form} \ge  0.6$ subsample within the $1\sigma$ uncertainty level (see Table~\ref{tab:plane300_zform} in Appendix~\ref{sec:tables}). This suggests that the dependence of the CFP on the formation epoch only comes into play for the most recently formed clusters with $z_\mathrm{form}<0.4$, or equivalently, those clusters formed over the last $\sim 4$~Gyr.

Consequently, there are hints pointing out that the CFP of recently formed clusters differs from that of clusters assembled at higher redshift. This conclusion holds irrespective of the definition and radial range of weighted temperature, at a significance level higher than $1\sigma$ in all cases. The typical differences between the CFP parameters of clusters with $z_\mathrm{form} < 0.4$ and $z_\mathrm{form}\ge 0.6$ are approximately $\Delta \alpha\sim 0.21 \pm 0.03$ and $\Delta \beta\sim -0.26 \pm 0.05$. The largest discrepancy between these two subsamples is found when the CFP is defined using the innermost region of clusters (i.e., $r\le 500h^{-1}$~kpc) to compute the weighted temperature, where the differences in the CFP parameters are of the order of $\Delta \alpha \sim 0.45 \pm 0.12$ and $\Delta \beta \sim -0.53 \pm 0.18$. Moreover, the CFP thickness of $z_\mathrm{form} < 0.4$ clusters is larger than for clusters formed at higher redshift. This reflects that recently formed clusters tend to have disturbed internal structures, which increases the dispersion of the CFP.



\section{Fundamental plane of CLASH clusters}\label{sec:fplaneclash}


Making use of the CLASH data sets, we have also explored the fundamental plane obtained from observations. In this study, we re-analysed the $20$ CLASH clusters of \citet{Fujita2018a,Fujita2018b} by adopting the methodology described in Section~\ref{sec:method}. For the CLASH sample, we used X-ray temperatures measured in the radial range $50$--$500 h^{-1}$~kpc (see Section~\ref{sec:clash}).

\begin{figure*}
\centering
    \resizebox{\textwidth}{!}{\includegraphics[clip=True,trim=0.45cm 0 0.5cm 0]{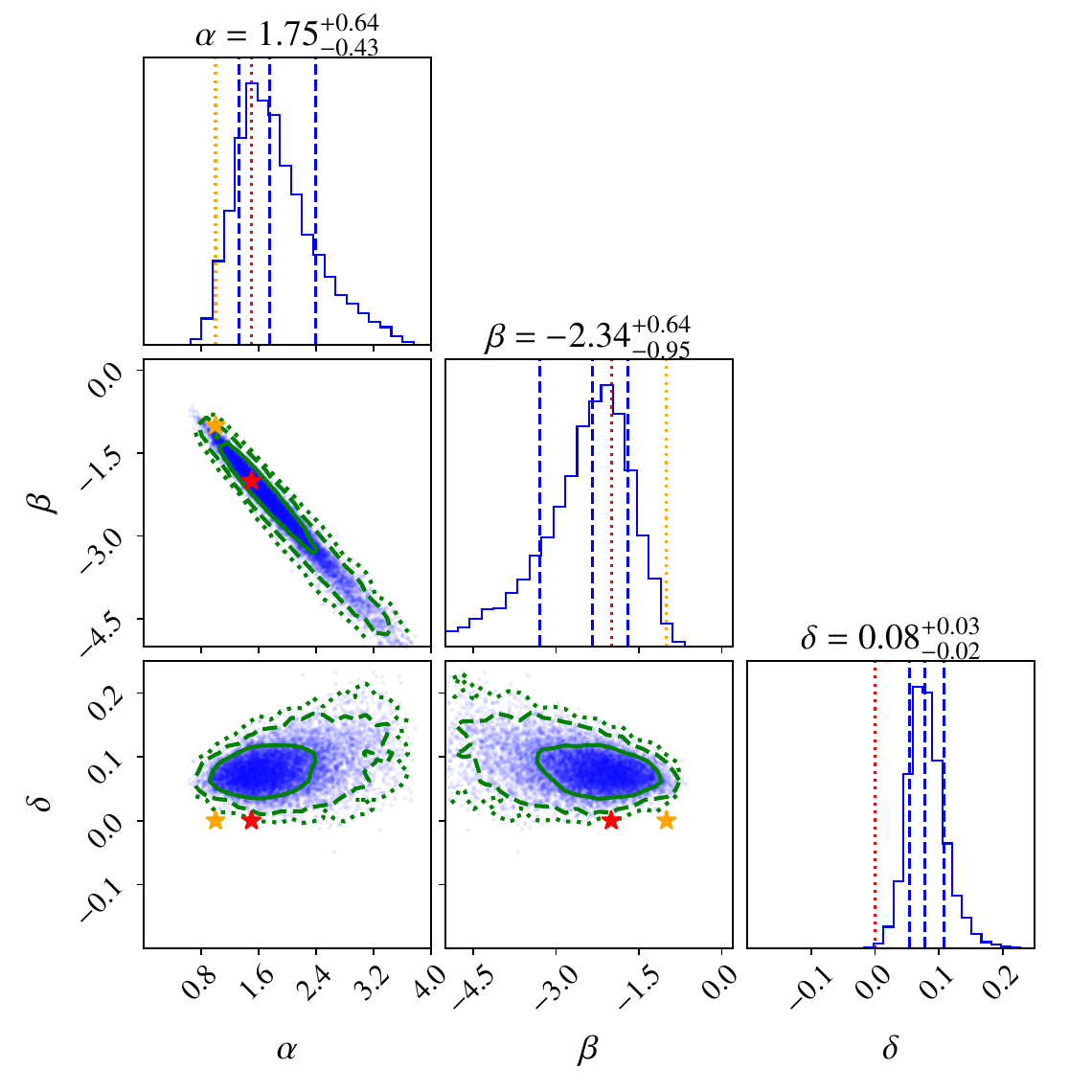}\includegraphics[clip=True,trim=0.45cm 0 1cm 0]{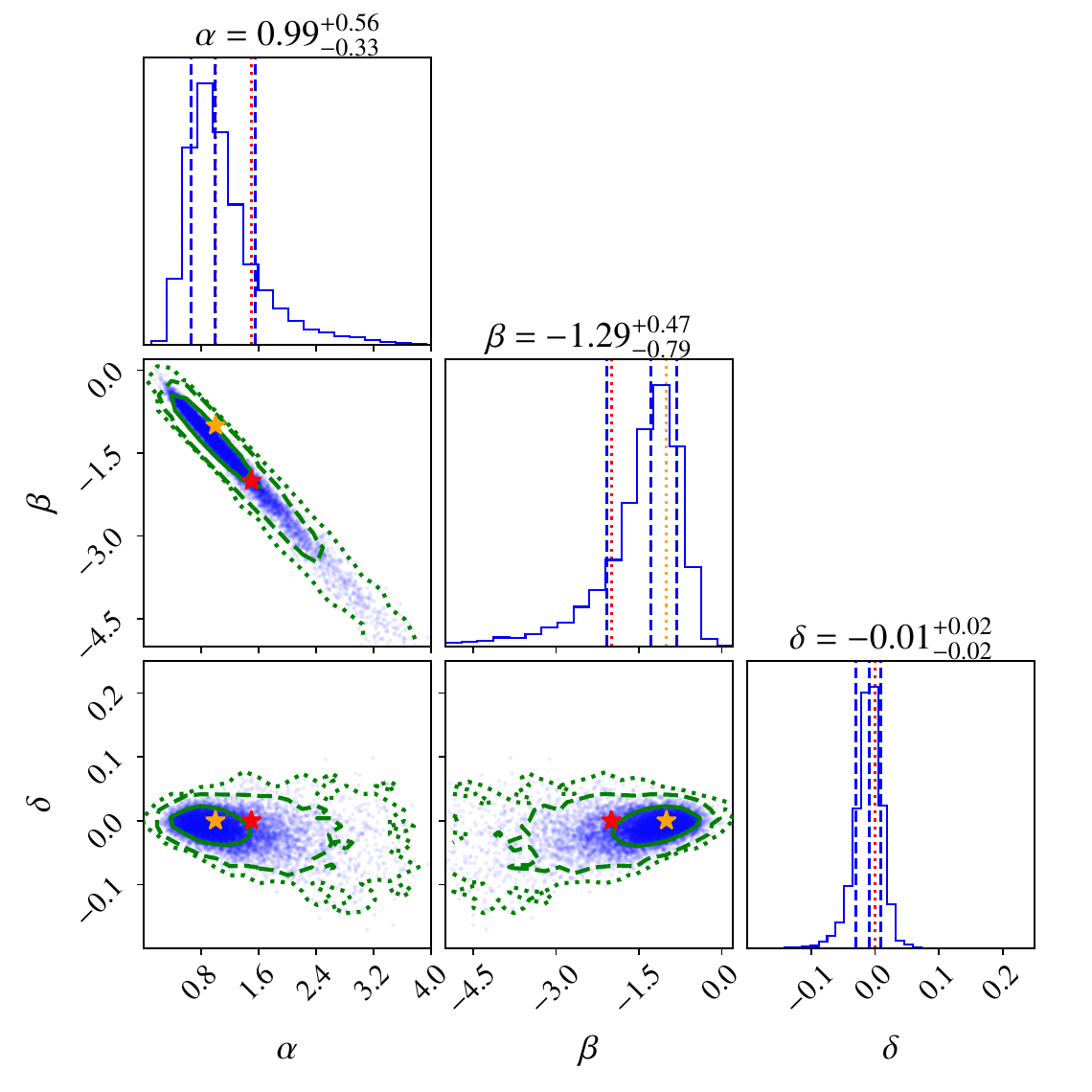}}
    \caption{Constraints on the parameters defining the fundamental plane (see equation~(\ref{eq:plane})) of CLASH galaxy clusters. Green solid, dashed, and dotted lines illustrate the confidence levels of $68$, $95$, and $99.7$ per cent of probability, respectively. Dashed blue lines show the $16^\mathrm{th}$, $50^\mathrm{th}$, and $84^\mathrm{th}$ percentiles of the distributions of parameters obtained from the posterior distributions of the NFW parameters for all individual clusters. The orange and red star-shaped symbols, as well as orange and red dotted lines in the histograms, indicate the virial expectation and the similarity solution for a secondary infall model, respectively.}
    \label{fig:mc_clash}
\end{figure*}

We find that the CLASH sample lies on a plane in logarithmic space defined by (\ms, \rs, and \tx), as previously found by \citet{Fujita2018b}. The best-fit parameters of $\alpha$, $\beta$, and $\delta$ obtained with MINDISQ (shown in Table~\ref{tab:plane_clash} and Fig.~\ref{fig:mc_clash}) provide the CFP with the lowest dispersion ($\sigma_\mathrm{d}=0.037$~dex) and these results are compatible with those obtained by \citet{Fujita2018b}:
\begin{equation}
    T \propto M_\mathrm{s}^{1.93^{+0.81}_{-0.57}}\ r_\mathrm{s}^{-2.57^{+0.83}_{-1.14}} \sim M_\mathrm{s}^{1.9}\ r_\mathrm{s}^{-2.6}
\end{equation}
We find that the discrepancies between the fundamental planes obtained for clusters of \three\ and those from CLASH are mainly due to the clusters with the highest X-ray temperatures. The rest of clusters lie on a shared region between both fundamental planes.

In addition, we explore whether the fundamental plane can be affected by the inclusion of unrelaxed clusters, even though the sample size is reduced. To this end, we removed those clusters identified as unrelaxed candidates in \citet[][see Table~\ref{tab:clash}]{Postman2012}. This results in a subsample of 13 CLASH clusters. For this `relaxed' subsample, we find that the resulting CFP is compatible with both the virial expectation and the similarity solution within the increased uncertainty, as shown in Table~\ref{tab:plane_clash}.  Instead, if we exclude the four hottest clusters with $T_\mathrm{X}>12$~keV (RX~J2248.7$-$4431, RX~J1347.5$-$1145. MACS~J0717.5$+$3745l, and MACS~J0647.7$+$7015), the CFP of the CLASH sample comes closer to the virial expectation, while the uncertainties of the CFP parameters decrease (see Table~\ref{tab:plane_clash} and right panel in Fig.~\ref{fig:mc_clash}). Hence, the four hottest clusters appear to be systematically deviated from the plane defined by the rest of the CLASH sample. 

The CFP obtained for a relaxed sample of \three\ clusters ($\chi_\mathrm{DS} \ge 1$) can be compared to the CLASH sample at a median redshift of $z\sim 0.35$. Using the spectroscopic-like temperature $T_\mathrm{sl}(500h^{-1}~\mathrm{kpc})$, the CFP parameters for the relaxed simulated sample at $z=0.35$ are obtained as $\alpha = 1.10 \pm 0.02$ and $\beta = -1.26 \pm 0.04$ (see Section~\ref{sec:relax_z} and Table~\ref{tab:coeff_z_log_relax}). This is in agreement with the CFP parameters ($\alpha,\beta$) obtained for a subsample of CLASH clusters with $T_\mathrm{X}$ lower than $12$~keV (see Table~\ref{tab:plane_clash}), above which no simulated clusters are found. Two of the four hottest CLASH clusters (MACS~J0717.5$+$3745 and MACS~J0647.7$+$7015) are high-magnification-selected systems at $z > 0.5$, which often turn out to be dynamically disturbed, highly massive ongoing mergers \citep[][]{Torri2004,Meneghetti2010,Meneghetti2014,Meneghetti2020,Umetsu2020rev}. The other two (RX~J2248.7$-$4431 and RX~J1347.5$-$1145) are X-ray-selected clusters that are classified as unrelaxed. The inclusion of the four clusters with $T_\mathrm{X} > 12$~keV alters the CFP for CLASH, in such a way that the CFP becomes compatible with the similarity solution. Here the main point is that these four clusters lie on the upper side of the CFP, so that they have a high contribution to determining the CFP for CLASH. This result suggests the possibility that the inclusion of clusters undergoing transient merger-induced boosts in temperature \citep{Ricker2001} can significantly affect the inference of the CFP parameters.

\begin{table*}
\centering
\caption{Best-fit parameters ($\alpha,\beta,\delta$) and thickness ($\sigma_\mathrm{d}$ in dex units) of the fundamental plane (see equation~(\ref{eq:plane})) derived for the CLASH sample. The upper and lower errors enclose the $1\sigma$ uncertainty range. The normalization parameters ($M_\mathrm{s,0}, r_\mathrm{s,0}, T_\mathrm{X,0}$) are also included in the table for each case.
\label{tab:plane_clash}}

\begin{tabular}{lcccccccc}
\hline
   & Number & $\alpha$ & $\beta$ & $\delta$ & $\sigma_\mathrm{d}$ & $M_\mathrm{s,0}$~ & $r_\mathrm{s,0}$ & $T_\mathrm{X,0}$\\
   & & & & & [dex] & [$10^{14}h^{-1}~M_\odot$] & [$h^{-1}$~kpc] & [keV] \\
\hline
CLASH         & $20$ & $1.75^{+0.64}_{-0.43}$ & $-2.34^{+0.64}_{-0.95}$ & $\ \ 0.078^{+0.030}_{-0.024}$ & $0.037$ &$2.96$ & $407$ & $7.75$\\
CLASH relaxed & $13$ & $1.82^{+0.86}_{-0.64}$ & $-2.45^{+0.96}_{-1.28}$ & $\ \ 0.135^{+0.056}_{-0.039}$ & $0.029$ & $2.38$ & $354$ & $6.70$\\
CLASH ($T_\mathrm{X} < 12~\mathrm{keV}$) & $16$ & $0.99^{+0.56}_{-0.33}$ & $-1.29^{+0.47}_{-0.79}$ & $-0.009^{+0.017}_{-0.021}$ & $0.032$ & $2.54$ & $399$ & $7.40$\\
 \hline
\end{tabular}
    
\end{table*}

\section{Discussion}\label{sec:discussion}


This section is devoted to a discussion of our findings from \three\ simulations. In Sections~\ref{sec:density_2mpc} and \ref{sec:sigmafit}, we explore potential sources of systematic effects that could bias our determination of the fundamental plane of galaxy clusters.  In Section~\ref{sec:other_simulations}, we analyse simulated cluster data in the published literature to derive their cluster fundamental planes and compare them with our results from \three\ simulations.


\subsection{Robustness of the NFW fitting procedure}\label{sec:density_2mpc}

Here we study systematic effects on the determination of the characteristic halo parameters \rs\ and \ms\ and their impact on the resulting CFP. Specifically, we investigate how the CFP parameters depend on the NFW fitting procedure using \three\ simulations.

To study possible effects of the choice of the fitting range, we repeat NFW fits using different radial ranges. For clusters at $z=0$, we refit the NFW formula to mass profiles $M(<r)$ of individual clusters in the radial range from $0.08\times r_{100}$ to $2h^{-1}$~Mpc. For clusters at higher redshifts ($z=0.07$, $0.22$, $0.33$, $0.59$, and $0.99$), we perform NFW fits in the radial range from $0.1\times r_{200}$ to $2h^{-1}$~Mpc. 

We also repeat NFW fits using the density profile $\rho(r)$, instead of the mass profile $M(<r)$. We fit the NFW model (equation~(\ref{eq:nfw})) to density profiles of individual clusters to determine \rs\ and \ms\ in two different radial ranges, namely: [$0.08, 1.0$]$\times r_{100}$ and from $0.08\times r_{100}$ to $2h^{-1}$~Mpc for clusters at $z=0$. For clusters at higher redshifts ($z=0.07$, $0.22$, $0.33$, $0.59$, and $0.99$), we perform NFW fits to density profiles in radial ranges of [$0.1, 1.0$]$\times r_{200}$ and from $0.1\times r_{200}$ to $2h^{-1}$~Mpc.

With new sets of \rs\ and \ms, we have repeated our CFP analysis, focusing on the dependence of the CFP on different temperature definitions, the redshift evolution of the CFP, and its dependence on the dynamical state of clusters.  As a result, we find no evidence of significant differences in the CFP parameters when using different NFW fitting procedures for extracting the \rs\ and \ms\ parameters. Therefore, our main conclusions remain unchanged. We only find mild discrepancies typically amounting to $|\Delta \alpha|$, $|\Delta \beta| \lesssim 0.1$.


\subsection{Robustness of the $\sigma_\mathrm{fit}$ selection}\label{sec:sigmafit}

To study the cluster fundamental plane, we constructed a simulated cluster sample from \three\ by selecting halos according to the fitting quality of the NFW profile, $\sigma_\mathrm{fit}=\chi^2/\mathrm{dof}\le 2$. This selection is to remove from the analysis those clusters that are not properly described by an NFW profile owing to the presence of strong asymmetries and massive substructures associated with mergers. It should be noted that, however, cosmological $N$-body simulations of $\Lambda$CDM reveal systematic deviations of quasi-equilibrium density profiles of collisionless halos from the self-similar NFW form \citep[e.g.,][]{Navarro2004,Merritt2006}. These numerical studies found that the density profiles of CDM halos are generally better fitted by a three-parameter Einasto profile, while both NFW and Einasto profiles describe well the density profiles of cluster-scale halos at low redshifts \citep[$z<1$;][]{Child2018}, as found by cluster lensing observations \citep[e.g.,][]{Umetsu2016}. Although the NFW assumption is well justified for our CFP analysis of cluster-scale CDM halos at $z<1$, we examine here the robustness of our findings as a function of the $\sigma_\mathrm{fit}$ threshold using \three\ simulations.

To this end, we repeat our CFP analysis for about $80$ values of the $\sigma_\mathrm{fit}$ threshold, ranging from $0.5$ up to $12$. About half of the \three\ clusters at $z=0$ have $\sigma_\mathrm{fit}$ values below $0.5$, so that a good fraction of our simulated clusters are properly fitted by an NFW profile. On the other hand, less than $15$~per cent of the clusters have $\sigma_\mathrm{fit}>3$, and less than  $8$~per cent present $\sigma_\mathrm{fit}>5$.

\begin{figure}
\centering
    \includegraphics[width=\columnwidth,clip=True]{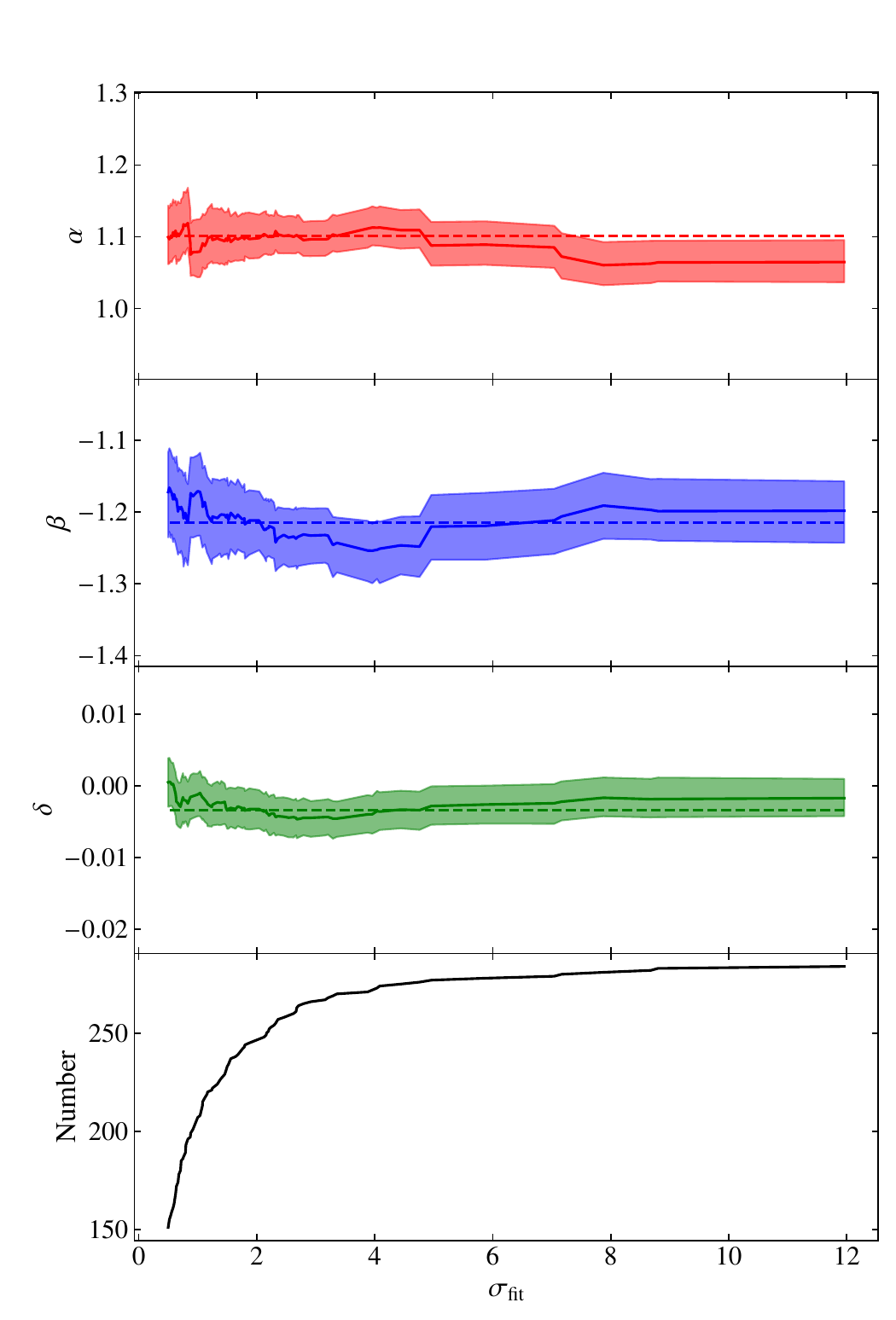}
    \caption{
    Best-fit parameters describing the fundamental plane at $z=0$ (see equation~(\ref{eq:plane})) as a function of the $\sigma_\mathrm{fit}$ threshold obtained for \three\ clusters using the mass-weighted temperature within $500h^{-1}$~kpc. \textit{From top to bottom}: $\alpha$, $\beta$, $\delta$, and number of clusters included in the analysis. In each panel, the shaded area shows the $1\sigma$ uncertainty range obtained by a bootstrapping technique. The dashed lines show the best-fit parameters for $\sigma_\mathrm{fit}\le2$.
    \label{fig:sigmalim_300}}
\end{figure}

Overall, we find that the resulting CFP parameters ($\alpha,\beta,\delta$) obtained for all cases at $z=0$ are independent of the $\sigma_\mathrm{fit}$ threshold and consistent within the $1\sigma$ uncertainty level, up to the $\sigma_\mathrm{fit}$ threshold of $6$.  Here the relative change in the CFP parameters is below $3$--$4$~per cent in all cases. This change in the parameters increases up to $8$~per cent, which corresponds to the selection threshold of $\sigma_\mathrm{fit}\le 12$ for the spectroscopic-like temperature within $500h^{-1}$~kpc.  In Fig.~\ref{fig:sigmalim_300}, we show the CFP parameters $\alpha$, $\beta$, and $\delta$ as a function of the $\sigma_\mathrm{fit}$ threshold.

We have also performed this test for the \three\ clusters at $z>0$, finding that the relative change in the CFP parameters ($\alpha,\beta,\delta$) decreases with increasing redshift. At $z=0.99$, the change in the CFP parameters is negligible regardless of the $\sigma_\mathrm{fit}$ threshold, even when including all clusters independently of the NFW fitting quality.

Finally, we also checked the stability of our results against the inclusion of clusters with extreme $\sigma_\mathrm{fit}$ values, namely, those with $\sigma_\mathrm{fit}\le 0.05$ and $\sigma_\mathrm{fit}\ge 2$. We find that these are homogeneously distributed across the range of the parameters explored in \three\ and they do not preferentially populate particular regions of the fundamental plane. We thus conclude that our simulation results do not depend on the particular choice of the $\sigma_\mathrm{fit}$ threshold.

\subsection{Fundamental planes from other cluster simulations}\label{sec:other_simulations}

The central gas properties in galaxy clusters and their connection with the dark-matter potential are sensitive to the effect of baryonic feedback. Hence, different implementations of baryonic physics in cosmological cluster simulations, as well as different cluster selection procedures, may lead to different predictions of the fundamental plane of galaxy clusters. In this subsection, we confront our findings with those obtained from the adiabatic MUSIC $N$-body/hydrodynamical simulations \citep[][Section~\ref{sec:music}]{Meneghetti2014} and another set of simulation data including non-gravitational feedback detailed in \citet[][hereafter FB simulations; Section~\ref{sec:FB}]{Rasia2015, Planelles2017}. It is of note that these simulations were also used to explore the CFP in \citet{Fujita2018b}.

\begin{table*}
\centering
\caption{Best-fit parameters ($\alpha,\beta,\delta$) and thickness ($\sigma_\mathrm{d}$ in dex units) of the fundamental plane obtained for simulated cluster samples, MUSIC at $z=0.25$, FB0 at $z=0$,  and FB1 at $z=1$. The lower and upper errors enclose the 1-$\sigma$ uncertainty.
\label{tab:planesim}}
    
\begin{tabular}{lcccccccc}
\hline
& Number & $\alpha$ & $\beta$ & $\delta$ & $\sigma_\mathrm{d}$ & $M_\mathrm{s,0}$ & $r_\mathrm{s,0}$ & $T_\mathrm{0}$ \\
& & & & & [dex] & [$10^{14}h^{-1}~M_\odot$] & [$h^{-1}$~kpc] & [keV] \\
\hline
MUSIC\ $T_\mathrm{mw}(r_{500})$ & $402$ & $1.28^{+0.03}_{-0.02}$ & $-1.55^{+0.05}_{-0.05}$ & \ \ $0.02^{+0.00}_{-0.00}$ & $0.023$ & $0.97$ & $283$ & $3.68$ \\
FB0\ $T_\mathrm{mw}(500h^{-1}\mathrm{kpc})$ & $29$ & $1.36^{+0.12}_{-0.10}$ & $-1.88^{+0.20}_{-0.25}$ & $-0.07^{+0.01}_{-0.01}$ & $0.027$ & $2.07$ & $444$ & $7.91$ \\
FB0\ $T_\mathrm{mw}(r_{500})$ & $29$ & $1.25^{+0.09}_{-0.08}$ & $-1.66^{+0.16}_{-0.20}$ & $-0.06^{+0.01}_{-0.01}$ & $0.022$ & $2.07$ & $444$ & $6.73$ \\
FB0\ $T_\mathrm{sl}(500h^{-1}\mathrm{kpc})$ & $29$ & $1.32^{+0.15}_{-0.13}$ & $-1.84^{+0.27}_{-0.32}$ & $-0.07^{+0.02}_{-0.02}$ & $0.030$ & $2.07$ & $444$ & $7.13$ \\
FB0\ $T_\mathrm{sl}(r_{500})$ & $29$ & $1.19^{+0.08}_{-0.06}$ & $-1.62^{+0.12}_{-0.15}$ & $-0.07^{+0.01}_{-0.01}$ & $0.016$ & $2.07$ & $444$ & $6.61$ \\
FB1\ $T_\mathrm{mw}(500h^{-1}\mathrm{kpc})$ & $29$ & $1.29^{+0.14}_{-0.12}$ & $-1.87^{+0.35}_{-0.34}$ & \ \ $0.02^{+0.02}_{-0.02}$ & $0.034$ & $0.47$ & $187$ & $3.27$ \\
FB1\ $T_\mathrm{sl}(500h^{-1}\mathrm{kpc})$ & $29$ & $1.17^{+0.13}_{-0.11}$ & $-1.69^{+0.34}_{-0.33}$ & \ \ $0.05^{+0.02}_{-0.02}$ & $0.030$ & $0.47$ & $187$ & $2.97$ \\
\hline
\end{tabular}
    
\end{table*}

\begin{figure}
\centering
\includegraphics[width=0.9\columnwidth,clip=True]{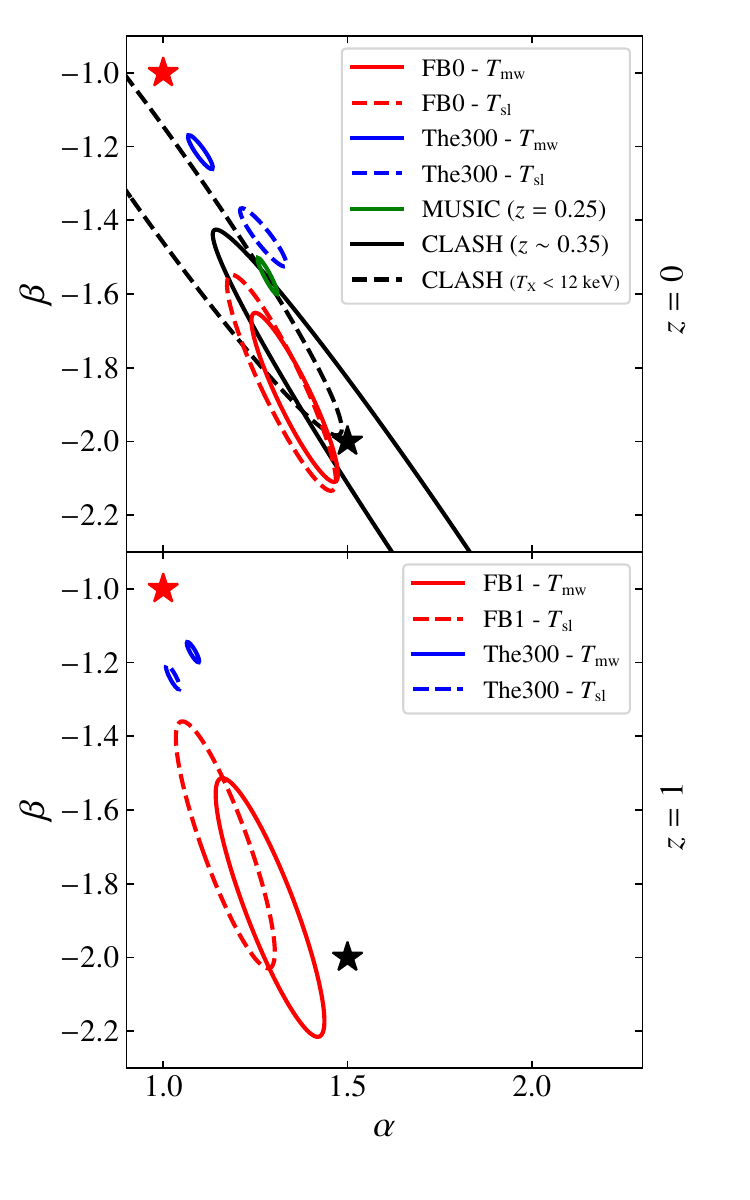}
\caption{
Constraints on the parameters $\alpha$ and $\beta$ defining the fundamental plane of galaxy clusters at $z=0$ obtained for \three\ and FB simulations (top panel, blue a red lines, respectively) using mass-weighted and spectroscopic-like temperatures measured within $500 h^{-1}$~kpc (solid and dashed lines, respectively). The contours show the $68$~per~cent confidence intervals in the $\alpha$--$\beta$ plane. The results obtained for the MUSIC simulations (green line) and CLASH (black lines), as well as the expectations for simplified virial equilibrium (red filled star) and the similarity solution for a secondary infall model (black filled star), are also shown. Similarly, the constraints on ($\alpha,\beta$) at $z=1$ for \three\ and FB simulations are shown in the bottom panel.}
\label{fig:comparison}
\end{figure}


\subsubsection{MUSIC simulations}\label{sec:music}

The MUSIC sample consists of resimulated halos selected from the dark-matter-only suit of MultiDark simulations ($\Omega_\mathrm{M}=0.27$, $\Omega_\mathrm{b}=0.0469$, $\Omega_\Lambda=0.73$, $\sigma_8=0.82$, $n_\mathrm{s}=0.95$, and $h=0.7$), whose halos are more massive than $10^{15}h^{-1}~M_\odot$ at $z=0$. For the high-resolution resimulation of Lagrangian regions around halos (a spherical region with radius of $6h^{-1}$~Mpc at $z=0$), the \textsc{TREEPM+SFH GADGET} code \citep[][]{Springel2005} was used, reaching a mass resolution for dark matter and gas of $m_\mathrm{DM}=9.01\times 10^{8}h^{-1}~M_\odot$ and $m_\mathrm{gas}=1.9\times10^8 h^{-1}~M_\odot$, respectively. The MUSIC simulations do not include any non-gravitational effects, such as AGN and supernova feedback, and there is no radiative cooling implemented.

Following \citet[][]{Fujita2018b}, we select all MUSIC clusters with $M_{200}>2\times 10^{14}h^{-1}~M_\odot$ at $z=0.25$ regardless of their dynamical state, amounting a total of $402$ clusters. Mass-weighted temperatures were computed for these clusters within the $r_{500}$ radial range, without excluding the core region \citep[][]{Fujita2018b}. This decision is based on that the MUSIC sample does not present cool-core features because the MUSIC simulations are non-radiative. The CFP of MUSIC clusters and the uncertainties in the CFP parameters are computed following the methodology described in Section~\ref{sec:method_eq}. For this sample, however, it is not possible to remove clusters that are not properly fitted by an NFW profile (i.e., no $\sigma_\mathrm{fit}$-selection applied) and the only temperature employed is $T_\mathrm{mw}(r_{500})$.

We find that the CFP of MUSIC clusters is characterised by $\alpha=1.28^{+0.03}_{-0.02}$ and  $\beta=-1.55^{+0.05}_{-0.05}$ (see Table~\ref{tab:planesim} and Fig.~\ref{fig:comparison}), which lie between the simplified virial expectation ($\alpha=1, \beta=-1$) and the similarity solution for a secondary infall model ($\alpha=1.5, \beta=-2$). In contrast, we find $\alpha=1.00^{+0.02}_{-0.02}$ and $\beta=-1.12^{+0.03}_{-0.03}$ for \three\ clusters at $z=0.22$ using the same temperature definition (see Table~\ref{tab:cfp_allz} in Appendix~\ref{sec:tables}), which is closer to the simplified virial expectation and significantly different from the MUSIC results at $z=0.25$. Moreover, the CFP obtained with the MUSIC simulations has a larger dispersion of $\sigma_\mathrm{d}=0.023$~dex, compared to $\sigma_\mathrm{d}=0.018$~dex for the \three\ simulations. The main difference between the MUSIC and \three\ simulations is the implementation of non-gravitational feedback (see Section~\ref{sec:300th}), meaning that both supernova feedback and AGN feedback were not implemented in the former set of simulations. This may suggest that quantitative predictions of the CFP depend on the implementation of non-gravitational physics to some degree.


\subsubsection{FB0 and FB1 simulations}\label{sec:FB} 

The FB simulation sample consists of $29$ massive halos with $M_{200} \in [1$--$30]\times 10^{14}h^{-1}~M_\odot$ at $z=0$, selected from a parent $N$-body cosmological simulation \citep[ $\Omega_\mathrm{M}=0.24$, $\Omega_\mathrm{b}= 0.04$, $\Omega_\Lambda=0.76$, $\sigma_8=0.8$, $n_\mathrm{s}=0.96$, and $h=0.72$; see][]{Rasia2015,Planelles2017}. Lagrangian regions around the selected clusters were resimulated with an improved resolution to include the baryonic component in hydrodynamic simulations. These simulations were carried out with the \textsc{GADGET} code, including an updated SPH scheme detailed in \citet{Beck2016}, reaching a final mass resolution of $m_\mathrm{DM} = 8.47\times 10^8 h^{-1}~M_\odot$ and $m_\mathrm{gas} = 1.53\times 10^8 h^{-1} ~M_\odot$ for the dark matter and gas components, respectively. The FB simulations include baryonic feedback effects, such as radiative cooling, star formation, supernova feedback, and metal enrichment, as well as AGN feedback \citep[for further details, see][]{Rasia2015,Planelles2017}. We note that the code used to perform the \three\ simulations is essentially the same as for the FB simulations, where the main differences are certain choices in stellar evolution and AGN feedback, as well as in the resimulated regions.

For our analysis, we select from \citet[][]{Fujita2018b} all $29$ clusters for the simulation runs at $z=0$ and $z=1$ (hereafter FB0 and FB1, respectively). Since this set of simulations includes radiative processes, mass-weighted and spectroscopic-like temperatures were computed within core-excised regions of $r\in [50,500]\times h^{-1}$~kpc and $[0.15,1.0]\times r_{500}$ \citep[][]{Fujita2018b}. For the FB1 run, both weighted temperatures are measured in the $[50,500]\times h^{-1}$~kpc radial range. We have repeated our CFP analysis on the FB cluster sample at $z=0$ and $z=1$ following the procedure described in Section~\ref{sec:method_eq}. It is worth mentioning that we used the same simulated sample than in \citet[][]{Fujita2018b} to explore the CFP by simulated clusters, where the only difference was the analysis outlined in Section~\ref{sec:method_eq}.

For the FB0 run, we find the best-fit CFP parameters in the range $\alpha \in [1.19, 1.36]$ and $\beta \in [-1.62, -1.88]$ (see Table~\ref{tab:planesim}). The best-fit $\alpha$ and $\beta$ parameters lie halfway between the simplified virial expectation and the similarity solution for a secondary infall model. These results are compatible with the similarity solution within a $95$~per~cent uncertainty level (see Fig.~\ref{fig:comparison}). We find no evidence of discrepancies between the results obtained using the spectroscopic-like and mass-weighted temperatures. We note that for the FB simulations, the uncertainties on the CFP parameters are typically larger owing to the small sample size of simulated clusters. For \three\ simulations, we find a tendency that the absolute values of $\alpha$ and $\beta$ decrease with increasing aperture radius within which the weighted temperature is computed (see Table~\ref{tab:planesim}).\footnote{However, we reiterate again that for the larger aperture radii of $r_{500}$ and $r_{200}$, there is a substantial aperture mismatch with respect to the characteristic scale radius $r_\mathrm{s}$ of halos.}  The absolute values of $\alpha$ and $\beta$ obtained for the FB0 sample are systematically larger than the corresponding \three\ results  (see Table~\ref{tab:plane300}).  Overall, the dispersion of the CFP for the FB0 sample is comparable to, but in some cases larger than, the corresponding \three\ results. 

For the FB1 run, we find a similar set of $\alpha$ and $\beta$ values to the FB0 results (Table~\ref{tab:planesim}). Comparing the FB0 and FB1 results, we find that there is a slight evolution of the CFP parameters, thus leading to similar conclusions. In particular, we notice that the CFP parameters ($\alpha,\beta$) for the FB1 sample are slightly less compatible with the similarity solution compared to the FB0 case. The variations in the CFP parameters are still compatible with no evolution given the large uncertainties. Nevertheless, the trends are qualitatively consistent with the \three\ results (see Section~\ref{sec:cfp_z}). Overall, the absolute values of $\alpha$ and $\beta$ increase with decreasing redshift, coming closer to the similarity solution.


\section{Summary and conclusions}\label{sec:summary}


Recent observational studies \citep{Fujita2018a,Fujita2018b} suggested that galaxy clusters form a tight fundamental plane in logarithmic space of the gas temperature ($T$) and the characteristic halo scale radius and mass ($r_\mathrm{s}$ and $M_\mathrm{s}$), that is, $T \propto M_\mathrm{s}^\alpha\,r_\mathrm{s}^\beta$. This cluster fundamental plane (CFP) was found to deviate from the virial equilibrium expectation, $T\propto M_\mathrm{s}\,r_\mathrm{s}^{-1}$, and to be in better agreement with the similarity solution for a secondary infall model \citep{Bertschinger1985}, $T\propto M_\mathrm{s}^{1.5}\,r_\mathrm{s}^{-2}$. 

In this paper, we have carried out a systematic study of the CFP using a sample of $\sim250$ simulated clusters from \three\ \citep{Cui2018}. In particular, we focus on the stability of the plane with different temperature definitions and its dependence on the dynamical relaxation state of clusters. The characteristic scale parameters of \three\ clusters, \rs\ and \ms, were extracted from the total mass profiles $M(<r)$ of individual clusters assuming an NFW halo description. After excluding clusters with poor fit quality of the NFW profile ($\sigma_\mathrm{fit}=\chi^2/\mathrm{dof} > 2$), we have $245$ clusters in our sample at $z=0$. At higher redshifts ($z=0.07$, $0.22$, $0.33$, $0.59$, and $0.99$), the number of selected clusters ranges from $231$ to $280$. We have explored two definitions of weighted temperatures, namely mass-weighted and spectroscopic-like temperatures ($T_\mathrm{mw}$ and $T_\mathrm{sl}$), and computed them in three radial ranges: $[0.1, 1.0]\times r_{200}$, $[0.15,1.0]\times r_{500}$, and $[50,500]\times h^{-1}$~kpc. 

We find that \three\ clusters at $z=0$ lie on a thin plane whose parameters ($\alpha$ and $\beta$) and dispersion ($\sigma_\mathrm{d}=0.015$--$0.030$~dex) depend on the gas temperature definition (Section~\ref{sec:fplane300}, Table~\ref{tab:plane300}, Figs.~\ref{fig:fplane_300} and \ref{fig:mc300}). Overall, the resulting CFP parameters are found in the range $1 < \alpha < 1.5$ and $-2 < \beta < -1$, that is, the range bounded by the virial equilibrium expectation and the similarity solution. The CFP for mass-weighted temperatures is slightly closer to the virial expectation ($\alpha=1$, $\beta=-1$) with a smaller dispersion, whereas the CFP parameters for the spectroscopic-like temperature lie halfway values between the virial expectation and the similarity solution. When gas temperatures are measured within $500h^{-1}$~kpc, which is close to the median value of \rs\ for the \three\ sample, the resulting CFP deviates the most from the virial expectation and shifts towards the similarity solution ($\alpha=1.5$, $\beta=-2$).

We have explored the evolution of the CFP with redshift for relaxed and unrelaxed clusters separately (Section~\ref{sec:relaxation}). Independently of the temperature definition, we find that clusters at $z=1$ form a CFP similar to the virial expectation, which evolves into a CFP that is halfway between the virial expectation and the similarity solution (Table~\ref{tab:coeff_z_log_relax}, Fig.~\ref{fig:coeff_z_relax}). For the unrelaxed subsample, the absolute values of $\alpha$ and $\beta$ progressively increase towards lower redshifts, coming closer to the similarity solution. In contrast, the CFP of relaxed clusters remains close to the virial expectation, with a milder evolution than for the unrelaxed subsample. Importantly, at all epochs, the CFP remains well defined throughout the evolution of the cluster population.

We have also studied the dependence of the CFP on the formation redshift, $z_\mathrm{form}$, using \three\ clusters at $z=0$ (Section~\ref{sec:zform}). Our results suggest that the CFP of recently formed clusters ($z_\mathrm{form}<0.4$) differs from that of clusters assembled at higher redshift, independently of the temperature definition (Table~\ref{tab:plane300_zform} in Appendix~\ref{sec:tables}). We find that clusters with $z_\mathrm{form}< 0.4$ form a CFP at $z=0$ that is most deviated from the virial expectation. In particular, the results obtained with the spectroscopic-like temperature within $500h^{-1}$~kpc are compatible with the similarity solution. These findings are in line with the results for the unrelaxed subsample. In fact, we verify that all \three\ clusters with $z_\mathrm{form} < 0.4$ are classified as unrelaxed.

Making use of the multiwavelength CLASH data sets, we also examined the fundamental plane for real clusters (Section~\ref{sec:fplaneclash}). We find that the CLASH sample forms a CFP with $\alpha=1.93^{+0.81}_{-0.57}$ and $\beta=-2.57^{+0.83}_{-1.14}$, with a dispersion of $\sigma_\mathrm{d}=0.037$~dex (see Table~\ref{tab:plane_clash} and left panel in Fig.~\ref{fig:mc_clash}). The results are compatible with those obtained by \citet{Fujita2018b} using the same data but with a different fitting procedure for the CFP measurement. There is a slight discrepancy between the fundamental planes obtained for clusters in \three\ and the CLASH sample (Fig.~\ref{fig:comparison}). We find that the inclusion of the four hottest clusters with $T_\mathrm{X}>12$~keV alters the CFP for CLASH. Excluding them, we find that the CFP for CLASH is fully compatible with the simulation results with \three\ clusters. This result suggests the possibility that the inclusion of clusters undergoing mergers \citep{Ricker2001,Rasia2011} can significantly affect the inference of the CFP parameters. 

Moreover, we explored potential sources of systematic effects that could potentially bias the determination of the CFP, such as the NFW fitting procedure and the effects of the $\sigma_\mathrm{fit}$ selection (Section~\ref{sec:discussion}). We conclude that all the results presented in this paper are robust against these effects. Moreover, we confront our results with the CFP obtained for other cosmological cluster simulations, finding that the overall trends are consistent with our findings from \three\ simulations.

This study has been centred on massive halos (all clusters having $T>2.5$~keV and the average temperature of the sample exceeding $5$~keV) which for the hierarchical structure formation model are the latest objects to form. As a result, the entire sample includes mostly unrelaxed clusters at all redshifts. Despite this condition, we always find that the CFP is extremely well defined with a low dispersion around the plane ($\sigma_\mathrm{d}<0.03$~dex). Even restricting the analysis to only the most disturbed systems or the most recently formed objects, the dispersion remains low.  We expect that the reduced dispersion not only holds but even improves when including lower mass systems that are not probed in this work, because smaller objects form earlier on average and are thus more relaxed. Thus far, the CFP has been directly measured only for the CLASH sample, which is quite unique for its characteristics and is also composed of extremely massive objects. Therefore, it would be interesting to extend the analysis to cluster samples with a wider mass range as targeted by ongoing cluster programs, such as the CHEX-MATE program \citep{CHEX-MATE2021}, the XXL X-ray survey \citep{Pierre2016}, and the eROSITA X-ray survey \citep{Brunner2021}, as well as blind Sunyaev--Zel'dovich effect surveys \citep[e.g.,][]{Hilton2021} and optical and near-infrared imaging surveys (e.g., Rubin Observatory LSST, Euclid, and Nancy Grace Roman Space Telescope missions).


\section*{Acknowledgements}


L.A.D.G.~and K.U.~acknowledge support from the Ministry of Science and Technology of Taiwan (grants MOST 106-2628-M-001-003-MY3 and MOST 109-2112-M-001-018-MY3) and from the Academia Sinica (grant AS-IA-107-M01). L.A.D.G.~thanks for financial support from the State Agency for Research of the Spanish MCIU through the `Center of Excellence Severo Ochoa' award to the Instituto de Astrof\'isica de Andaluc\'ia (SEV-2017-0709), and to the PID2019-109067-GB100. E.L.~receives support by contract ASI-INAF n.2017-14-H.0, whereas W.C.~is supported by the European Research Council under grant number 670193 and by the STFC AGP Grant ST/V000594/1 and further acknowledges the science research grants from the China Manned Space Project with No.~CMS-CSST-2021-A01 and CMS-CSST-2021-B01. The authors are also grateful to Yutaka Fujita for sharing part of tables employed in his work, as well as the discussions carried out with him during this research, which contributed to improve the present work. In addition, the authors thank the constructive and fruitful comments provided by the referee that helped to improve the content of this paper.

This work has been made possible by The Three Hundred Collaboration. The simulations used in this paper have been performed in the MareNostrum Supercomputer at the Barcelona Supercomputing Center, thanks to CPU time granted by the Red Espa\~nola de Supercomputaci\'on. As part of The Three Hundred Project, this work has received financial support from the European Union's Horizon 2020 Research and Innovation Programme under the Marie Sklodowska-Curie grant agreement number 734374, the LACEGAL project. For this research, we made use of multiple \textsc{Python} packages, such as: \textsc{Matplotlib} \citep[][]{Hunter2007}, \textsc{Corner} \citep{corner}, and \textsc{Numpy} \citep[][]{Harris2020}.


\section*{Data Availability}

The data used in this paper are part of \three\ collaboration and can be accessed following the guidelines in the main website\footnote{https://the300-project.org} of the project.




\bibliographystyle{mnras}
\bibliography{fplane_v2.1} 




\appendix

\section{Parameters defining the fundamental plane of Galaxy clusters}\label{sec:tables}

In this appendix, we summarise the parameters defining the CFP and its thickness for \three\ clusters determined at different redshifts and with different temperature definitions (Table~\ref{tab:cfp_allz}), and those obtained for the relaxed and unrelaxed subsamples at $z=0$ (Table~\ref{tab:plane300_relax}), and those at $z=0$ for subsamples with different formation redshifts (Table~\ref{tab:plane300_zform}).

\begin{table*}
\centering
\caption{Best-fit parameters ($\alpha,\beta,\delta$) and thickness ($\sigma_\mathrm{d}$ in dex units) of the fundamental plane obtained for \three\ clusters at different redshifts and with different temperature definitions. The results are obtained from NFW fits to the center-excised mass profile inside $r_{200}$. For each redshift, the number of clusters and the median values of the characteristic scale parameters, $r_\mathrm{s,0}$ and $M_\mathrm{s,0}$, are listed at the bottom of each panel. For each quantity, the upper and lower errors enclose the $1\sigma$ uncertainty. The dashes correspond to $\delta$ errors smaller than $5\times10^{-4}$.
\label{tab:cfp_allz}}

\begin{tabular}{lccccc}
\hline
$z=0.07$ & $\alpha$ & $\beta$ & $\delta$ & $\sigma_\mathrm{d}$ & $T_0$ \\
\hline
$T_\mathrm{mw}(500h^{-1}~\mathrm{kpc})$ & $1.18^{+0.03}_{-0.03}$ & $-1.30^{+0.04}_{-0.04}$ & $0.000^{--}_{--}$ & $0.017$ & $6.35$ \\
$T_\mathrm{mw}(r_{500})$ & $1.06^{+0.02}_{-0.02}$ & $-1.18^{+0.03}_{-0.03}$ & $0.000^{--}_{--}$ & $0.016$ & $5.67$ \\
$T_\mathrm{mw}(r_{200})$ & $1.03^{+0.02}_{-0.02}$ & $-1.13^{+0.03}_{-0.03}$ & $0.000^{--}_{--}$ & $0.015$ & $5.05$ \\
$T_\mathrm{sl}(500h^{-1}~\mathrm{kpc})$ & $1.23^{+0.05}_{-0.04}$ & $-1.47^{+0.06}_{-0.07}$ & $0.000^{--}_{--}$ & $0.022$ & $5.97$ \\
$T_\mathrm{sl}(r_{500})$ & $1.21^{+0.04}_{-0.03}$ & $-1.52^{+0.05}_{-0.06}$ & $0.016^{+0.003}_{-0.003}$ & $0.019$ & $5.54$ \\
$T_\mathrm{sl}(r_{200})$ & $1.23^{+0.04}_{-0.03}$ & $-1.56^{+0.05}_{-0.05}$ & $0.014^{+0.003}_{-0.003}$ & $0.018$ & $5.32$ \\
\hline
\multicolumn{2}{c}{Number=$231$} & \multicolumn{2}{c}{$M_\mathrm{s,0}=1.65 \times 10^{14}h^{-1}~M_\odot$} & \multicolumn{2}{c}{$r_\mathrm{s,0}=329h^{-1}$~kpc} \\
\end{tabular}\\
\begin{tabular}{lccccc}
\hline
$z=0.22$ & $\alpha$ & $\beta$ & $\delta$ & $\sigma_\mathrm{d}$ & $T_0$ \\
\hline
$T_\mathrm{mw}(500h^{-1}~\mathrm{kpc})$ & $1.10^{+0.02}_{-0.02}$ & $-1.23^{+0.03}_{-0.03}$ & $0.000^{--}_{--}$ & $0.016$ & $5.45$ \\
$T_\mathrm{mw}(r_{500})$ & $1.00^{+0.02}_{-0.02}$ & $-1.12^{+0.03}_{-0.03}$ & $0.007^{+0.002}_{-0.002}$ & $0.018$ & $4.93$ \\
$T_\mathrm{mw}(r_{200})$ & $0.97^{+0.02}_{-0.02}$ & $-1.06^{+0.03}_{-0.03}$ & $0.000^{--}_{--}$ & $0.018$ & $4.44$ \\
$T_\mathrm{sl}(500h^{-1}~\mathrm{kpc})$ & $1.13^{+0.03}_{-0.03}$ & $-1.35^{+0.04}_{-0.05}$ & $0.000^{--}_{--}$ & $0.020$ & $5.13$ \\
$T_\mathrm{sl}(r_{500})$ & $1.08^{+0.03}_{-0.02}$ & $-1.34^{+0.04}_{-0.04}$ & $0.000^{--}_{--}$ & $0.018$ & $4.86$ \\
$T_\mathrm{sl}(r_{200})$ & $1.12^{+0.03}_{-0.02}$ & $-1.41^{+0.04}_{-0.04}$ & $0.002^{+0.003}_{-0.003}$ & $0.019$ & $4.70$ \\
\hline
\multicolumn{2}{c}{Number=$243$} & \multicolumn{2}{c}{$M_\mathrm{s,0}=1.34 \times 10^{14}h^{-1}~M_\odot$} & \multicolumn{2}{c}{$r_\mathrm{s,0}=303h^{-1}$~kpc} \\
\end{tabular}\\
\begin{tabular}{lccccc}
\hline
$z=0.33$ & $\alpha$ & $\beta$ & $\delta$ & $\sigma_\mathrm{d}$ & $T_0$ \\
\hline
$T_\mathrm{mw}(500h^{-1}~\mathrm{kpc})$ & $1.10^{+0.02}_{-0.02}$ & $-1.26^{+0.03}_{-0.03}$ & $-0.005^{+0.002}_{-0.002}$ & $0.015$ & $5.09$ \\
$T_\mathrm{mw}(r_{500})$ & $1.00^{+0.01}_{-0.02}$ & $-1.13^{+0.03}_{-0.02}$ & $-0.011^{+0.002}_{-0.002}$ & $0.017$ & $4.73$ \\
$T_\mathrm{mw}(r_{200})$ & $0.98^{+0.02}_{-0.02}$ & $-1.07^{+0.03}_{-0.03}$ & $-0.006^{+0.002}_{-0.002}$ & $0.018$ & $4.18$ \\
$T_\mathrm{sl}(500h^{-1}~\mathrm{kpc})$ & $1.11^{+0.02}_{-0.02}$ & $-1.37^{+0.04}_{-0.04}$ & $\ \ 0.001^{+0.003}_{-0.003}$ & $0.019$ & $4.79$ \\
$T_\mathrm{sl}(r_{500})$ & $1.06^{+0.02}_{-0.02}$ & $-1.32^{+0.03}_{-0.04}$ & $-0.009^{+0.002}_{-0.002}$ & $0.018$ & $4.65$ \\
$T_\mathrm{sl}(r_{200})$ & $1.09^{+0.02}_{-0.02}$ & $-1.36^{+0.04}_{-0.04}$ & $\ \ 0.000^{+0.000}_{-0.000}$ & $0.018$ & $4.49$ \\
\hline
\multicolumn{2}{c}{Number=$243$} & \multicolumn{2}{c}{$M_\mathrm{s,0}=1.18 \times 10^{14}h^{-1}~M_\odot$} & \multicolumn{2}{c}{$r_\mathrm{s,0}=288h^{-1}$~kpc} \\
\end{tabular}\\
\begin{tabular}{lccccc}
\hline
$z=0.59$ & $\alpha$ & $\beta$ & $\delta$ & $\sigma_\mathrm{d}$ & $T_0$ \\
\hline
$T_\mathrm{mw}(500h^{-1}~\mathrm{kpc})$ & $1.10^{+0.02}_{-0.02}$ & $-1.23^{+0.03}_{-0.03}$ & $0.000^{--}_{--}$ & $0.017$ & $4.15$ \\
$T_\mathrm{mw}(r_{500})$ & $0.98^{+0.02}_{-0.02}$ & $-1.10^{+0.03}_{-0.03}$ & $0.000^{--}_{--}$ & $0.019$ & $3.96$ \\
$T_\mathrm{mw}(r_{200})$ & $0.95^{+0.02}_{-0.02}$ & $-1.04^{+0.03}_{-0.03}$ & $0.000^{--}_{--}$ & $0.019$ & $3.54$ \\
$T_\mathrm{sl}(500h^{-1}~\mathrm{kpc})$ & $1.09^{+0.02}_{-0.02}$ & $-1.34^{+0.04}_{-0.04}$ & $0.000^{--}_{--}$ & $0.019$ & $4.04$ \\
$T_\mathrm{sl}(r_{500})$ & $1.03^{+0.02}_{-0.02}$ & $-1.28^{+0.04}_{-0.03}$ & $0.000^{--}_{--}$ & $0.018$ & $3.90$ \\
$T_\mathrm{sl}(r_{200})$ & $1.05^{+0.02}_{-0.02}$ & $-1.31^{+0.04}_{-0.04}$ & $0.006^{+0.002}_{-0.002}$ & $0.019$ & $3.77$ \\
\hline
\multicolumn{2}{c}{Number=$274$} & \multicolumn{2}{c}{$M_\mathrm{s,0}=8.54 \times 10^{13}h^{-1}~M_\odot$} & \multicolumn{2}{c}{$r_\mathrm{s,0}=239h^{-1}$~kpc} \\
\end{tabular}\\
\begin{tabular}{lccccc}
\hline
$z=0.99$ & $\alpha$ & $\beta$ & $\delta$ & $\sigma_\mathrm{d}$ & $T_0$ \\
\hline
$T_\mathrm{mw}(500h^{-1}~\mathrm{kpc})$ & $1.08^{+0.01}_{-0.02}$ & $-1.17^{+0.03}_{-0.03}$ & $0.000^{--}_{--}$ & $0.017$ & $2.93$ \\
$T_\mathrm{mw}(r_{500})$ & $0.94^{+0.01}_{-0.02}$ & $-1.04^{+0.03}_{-0.03}$ & $0.000^{--}_{--}$ & $0.016$ & $3.10$ \\
$T_\mathrm{mw}(r_{200})$ & $0.94^{+0.02}_{-0.01}$ & $-1.01^{+0.03}_{-0.03}$ & $0.000^{--}_{--}$ & $0.016$ & $2.72$ \\
$T_\mathrm{sl}(500h^{-1}~\mathrm{kpc})$ & $1.03^{+0.02}_{-0.02}$ & $-1.24^{+0.03}_{-0.03}$ & $0.000^{--}_{--}$ & $0.017$ & $3.06$ \\
$T_\mathrm{sl}(r_{500})$ & $0.95^{+0.02}_{-0.02}$ & $-1.15^{+0.03}_{-0.03}$ & $0.000^{--}_{--}$ & $0.019$ & $3.10$ \\
$T_\mathrm{sl}(r_{200})$ & $0.97^{+0.02}_{-0.02}$ & $-1.19^{+0.04}_{-0.03}$ & $0.000^{+0.002}_{-0.002}$ & $0.018$ & $2.98$ \\
\hline
\multicolumn{2}{c}{Number=$280$} & \multicolumn{2}{c}{$M_\mathrm{s,0}=5.19 \times 10^{13}h^{-1}~M_\odot$} & \multicolumn{2}{c}{$r_\mathrm{s,0}=185h^{-1}$~kpc} \\
\hline
\end{tabular}

\end{table*}

\begin{table*}
\centering
\caption{Same as Table~\ref{tab:plane300}, but for the relaxed ($\relaxmath \ge 1$, upper panel) and unrelaxed ($\relaxmath \le 2/3$, lower panel) subsamples at $z=0$. \label{tab:plane300_relax}}

\begin{tabular}{lccccc}
\hline
 & $\alpha$ & $\beta$ & $\delta$ & $\sigma_\mathrm{d}$ & $T_\mathrm{0}$\\
 & & & & [dex] & [keV] \\
\hline
$T_\mathrm{mw}(500h^{-1}~\mathrm{kpc})$ & $1.01^{+0.05}_{-0.04}$ & $-0.99^{+0.07}_{-0.08}$ &  $-0.003^{+0.003}_{-0.003}$ & $0.018$ & $6.71$\\
$T_\mathrm{mw}(r_{500})$         & $0.96^{+0.04}_{-0.03}$ & $-1.02^{+0.05}_{-0.06}$ & $\ \ 0.001^{+0.002}_{-0.002}$ & $0.013$ & $5.53$\\
$T_\mathrm{mw}(r_{200})$         & $0.91^{+0.04}_{-0.03}$ & $-0.94^{+0.04}_{-0.05}$ & $\ \ 0.005^{+0.002}_{-0.002}$ & $0.012$ & $4.94$\\
$T_\mathrm{sl}(500h^{-1}~\mathrm{kpc})$ & $1.07^{+0.08}_{-0.05}$ & $-1.15^{+0.10}_{-0.11}$ & $-0.009^{+0.005}_{-0.005}$ & $0.026$ & $6.53$\\
$T_\mathrm{sl}(r_{500})$           & $1.05^{+0.06}_{-0.04}$ & $-1.23^{+0.07}_{-0.08}$ & $-0.007^{+0.003}_{-0.004}$ & $0.016$ & $5.80$\\
$T_\mathrm{sl}(r_{200})$           & $1.06^{+0.06}_{-0.04}$ & $-1.27^{+0.07}_{-0.08}$ & $-0.013^{+0.003}_{-0.004}$ & $0.015$ & $5.68$\\
\hline
\multicolumn{2}{c}{Number=$90$}  & \multicolumn{2}{c}{$M_\mathrm{s,0}=1.67 \times 10^{14}h^{-1}~M_\odot$} & \multicolumn{2}{c}{$r_\mathrm{s,0}=355h^{-1}$~kpc} \\
\hline
\end{tabular}

\begin{tabular}{lccccc}
\hline
 & $\alpha$ & $\beta$ & $\delta$ & $\sigma_\mathrm{d}$ & $T_\mathrm{0}$\\
 & & & & [dex] & [keV] \\
\hline
$T_\mathrm{mw}(500h^{-1}~\mathrm{kpc})$ & $1.20^{+0.13}_{-0.09}$ & $-1.50^{+0.11}_{-0.12}$ &  $-0.004^{+0.007}_{-0.007}$ & $0.022$ & $6.63$\\
$T_\mathrm{mw}(r_{500})$         & $1.02^{+0.07}_{-0.05}$ & $-1.15^{+0.07}_{-0.07}$ & $-0.008^{+0.004}_{-0.004}$ & $0.019$ & $5.71$\\
$T_\mathrm{mw}(r_{200})$         & $1.00^{+0.05}_{-0.03}$ & $-1.08^{+0.06}_{-0.06}$ &  $-0.006^{+0.004}_{-0.004}$ & $0.024$ & $5.09$\\
$T_\mathrm{sl}(500h^{-1}~\mathrm{kpc})$ & $1.40^{+0.22}_{-0.15}$ & $-1.71^{+0.16}_{-0.18}$ & $-0.019^{+0.009}_{-0.010}$ & $0.031$ & $6.04$\\
$T_\mathrm{sl}(r_{500})$           & $1.16^{+0.14}_{-0.09}$ & $-1.44^{+0.12}_{-0.12}$ & $-0.002^{+0.006}_{-0.007}$ & $0.023$ & $5.25$\\
$T_\mathrm{sl}(r_{200})$           & $1.20^{+0.13}_{-0.08}$ & $-1.50^{+0.11}_{-0.12}$ & $-0.004^{+0.007}_{-0.007}$ & $0.026$ & $4.96$\\
\hline
\multicolumn{2}{c}{Number=$87$}  & \multicolumn{2}{c}{$M_\mathrm{s,0}=2.49 \times 10^{14}h^{-1}~M_\odot$} & \multicolumn{2}{c}{$r_\mathrm{s,0}=513h^{-1}$~kpc} \\
\hline
\end{tabular}

\end{table*}

\begin{table*}
\centering
\caption{Same as Table~\ref{tab:plane300}, but for \three\ clusters at $z=0$ with different formation redshifts: $z_\mathrm{form} < 0.4$, $0.4 \le z_\mathrm{form} < 0.6$, and $z_\mathrm{form} \ge 0.6$.\label{tab:plane300_zform}}
    
\begin{tabular}{lccccc}
\hline
$z_\mathrm{form} \ge 0.6$  & $\alpha$ & $\beta$ & $\delta$ & $\sigma_\mathrm{d}$ & $T_\mathrm{0}$\\
 & & & & [dex] & [keV] \\
\hline
$T_\mathrm{mw}(500h^{-1}~\mathrm{kpc})$ & $1.04^{+0.04}_{-0.03}$ & $-1.09^{+0.06}_{-0.07}$ & $0.013^{+0.004}_{-0.004}$ & $0.021$ & $6.71$\\
$T_\mathrm{mw}(r_{500})$         & $0.98^{+0.03}_{-0.03}$ & $-1.05^{+0.05}_{-0.05}$ & $0.017^{+0.002}_{-0.002}$ & $0.012$ & $5.68$\\
$T_\mathrm{mw}(r_{200})$         & $0.92^{+0.04}_{-0.03}$ & $-0.94^{+0.05}_{-0.05}$ & $0.018^{+0.002}_{-0.002}$ & $0.012$ & $5.07$\\
$T_\mathrm{sl}(500h^{-1}~\mathrm{kpc})$ & $1.11^{+0.07}_{-0.05}$ & $-1.27^{+0.10}_{-0.11}$ & $0.018^{+0.006}_{-0.006}$ & $0.022$ & $6.42$\\
$T_\mathrm{sl}(r_{500})$           & $1.10^{+0.05}_{-0.04}$ & $-1.33^{+0.07}_{-0.08}$ & $0.019^{+0.004}_{-0.004}$ & $0.017$ & $5.82$\\
$T_\mathrm{sl}(r_{200})$           & $1.10^{+0.05}_{-0.04}$ & $-1.37^{+0.07}_{-0.09}$ & $0.013^{+0.004}_{-0.004}$ & $0.017$ & $5.69$\\
\hline
\multicolumn{2}{c}{Number=$87$}  & \multicolumn{2}{c}{$M_\mathrm{s,0}=1.66 \times 10^{14}h^{-1}~M_\odot$} & \multicolumn{2}{c}{$r_\mathrm{s,0}=336h^{-1}$~kpc} \\
\hline
\end{tabular}
    
\begin{tabular}{lccccc}
\hline
$0.4 \le z_\mathrm{form} < 0.6$  & $\alpha$ & $\beta$ & $\delta$ & $\sigma_\mathrm{d}$ & $T_\mathrm{0}$\\
 & & & & [dex] & [keV] \\
\hline
$T_\mathrm{mw}(500h^{-1}~\mathrm{kpc})$ & $1.03^{+0.07}_{-0.05}$ & $-1.08^{+0.08}_{-0.09}$ & $-0.021^{+0.005}_{-0.004}$ & $0.019$ & $6.97$\\
$T_\mathrm{mw}(r_{500})$         & $0.91^{+0.03}_{-0.03}$ & $-0.91^{+0.05}_{-0.06}$ & $-0.002^{+0.003}_{-0.003}$ & $0.013$ & $5.58$\\
$T_\mathrm{mw}(r_{200})$         & $0.91^{+0.02}_{-0.02}$ & $-0.89^{+0.05}_{-0.05}$ & $\ \ 0.000^{+0.003}_{-0.003}$ & $0.012$ & $4.95$\\
$T_\mathrm{sl}(500h^{-1}~\mathrm{kpc})$ & $1.14^{+0.11}_{-0.08}$ & $-1.41^{+0.13}_{-0.15}$ & $-0.020^{+0.007}_{-0.008}$ & $0.028$ & $6.42$\\
$T_\mathrm{sl}(r_{500})$           & $0.99^{+0.06}_{-0.04}$ & $-1.15^{+0.07}_{-0.07}$ & $\ \ 0.000^{+0.004}_{-0.004}$ & $0.016$ & $5.48$\\
$T_\mathrm{sl}(r_{200})$           & $1.06^{+0.07}_{-0.05}$ & $-1.32^{+0.10}_{-0.11}$ & $\ \ 0.000^{+0.005}_{-0.006}$ & $0.017$ & $5.16$\\
\hline
\multicolumn{2}{c}{Number=$75$}  & \multicolumn{2}{c}{$M_\mathrm{s,0}=2.02 \times 10^{14}h^{-1}~M_\odot$} & \multicolumn{2}{c}{$r_\mathrm{s,0}=431h^{-1}$~kpc} \\
\hline
\end{tabular}
    
\begin{tabular}{lccccc}
\hline
$z_\mathrm{form} < 0.4$ & $\alpha$ & $\beta$ & $\delta$ & $\sigma_\mathrm{d}$ & $T_\mathrm{0}$\\
 & & & & [dex] & [keV] \\
\hline
$T_\mathrm{mw}(500h^{-1}~\mathrm{kpc})$ & $1.27^{+0.06}_{-0.05}$ & $-1.46^{+0.07}_{-0.08}$ & $-0.002^{+0.005}_{-0.005}$ & $0.021$ & $6.55$\\
$T_\mathrm{mw}(r_{500})$         & $1.11^{+0.04}_{-0.04}$ & $-1.21^{+0.06}_{-0.06}$ & $-0.016^{+0.004}_{-0.004}$ & $0.017$ & $5.61$\\
$T_\mathrm{mw}(r_{200})$         & $1.06^{+0.04}_{-0.03}$ & $-1.12^{+0.06}_{-0.06}$ & $-0.016^{+0.004}_{-0.004}$ & $0.020$ & $5.07$\\
$T_\mathrm{sl}(500h^{-1}~\mathrm{kpc})$ & $1.55^{+0.13}_{-0.10}$ & $-1.80^{+0.13}_{-0.16}$ & $-0.029^{+0.009}_{-0.009}$ & $0.034$ & $6.03$\\
$T_\mathrm{sl}(r_{500})$           & $1.26^{+0.07}_{-0.06}$ & $-1.49^{+0.08}_{-0.08}$ & $-0.007^{+0.006}_{-0.006}$ & $0.020$ & $5.16$\\
$T_\mathrm{sl}(r_{200})$           & $1.25^{+0.06}_{-0.05}$ & $-1.51^{+0.07}_{-0.08}$ & $-0.016^{+0.005}_{-0.005}$ & $0.020$ & $5.01$\\
\hline
\multicolumn{2}{c}{Number=$83$}  & \multicolumn{2}{c}{$M_\mathrm{s,0}=2.49 \times 10^{14}h^{-1}~M_\odot$} & \multicolumn{2}{c}{$r_\mathrm{s,0}=518h^{-1}$~kpc} \\
\hline
\end{tabular}
    
\end{table*}



\bsp	
\label{lastpage}
\end{document}